\documentclass[11pt]{article}
\usepackage[hidelinks]{hyperref}
\usepackage{tikz}
\usetikzlibrary{calc}
\usepackage{multicol}
\usepackage{mdframed}
\usepackage{color}
\usepackage{float}
\usepackage{epsfig,subfigure}
\usepackage{kbordermatrix}
\usepackage{amssymb, amsmath}
\usepackage{color}
\usepackage{bm}
\usepackage{amsthm}
\usepackage{authblk}
\usepackage{cite}
\usetikzlibrary{shapes,decorations,arrows,calc,arrows.meta,fit,positioning}

\usepackage{graphicx}
\usepackage{color}

\usepackage{amssymb}
\usepackage{amsmath,amsfonts,amssymb}
\usepackage{verbatim}
\usepackage{stfloats}
\usepackage[bookmarks=false]{}

\newtheorem{theorem}{Theorem}

\newtheorem{lemma}{Lemma}

\newcommand{\vect}[1]{\overrightarrow{#1}}
\newcommand{\emphv}[1]{
\begin{tikzpicture}[baseline={([yshift=-0.5ex]current bounding box.center)}]
\node [draw, rounded corners]  {{$#1$}}; 
\end{tikzpicture}}

\begin{document}

\title{$X$-secure $T$-private Information Retrieval from MDS Coded Storage with Byzantine and Unresponsive Servers}
\author{Zhuqing Jia and Syed A. Jafar}
\affil{Center for Pervasive Communications and Computing (CPCC), UC Irvine\\
Email: \{zhuqingj, syed\}@uci.edu}
\date{}
\maketitle

\begin{abstract}
The problem of $X$-secure $T$-private information retrieval from MDS coded storage is studied in this paper, where the user wishes to privately retrieve one out of $K$ independent messages  that are distributed over $N$ servers according to an MDS code. It is guaranteed that any group of up to $X$ colluding servers learn nothing about the messages and that any group of up to $T$ colluding servers learn nothing about the identity of desired message. A lower bound of achievable rates is proved by presenting a novel scheme based on \emph{cross-subspace alignment} and a successive decoding with interference cancellation strategy. For large number of messages $(K\rightarrow\infty)$ the achieved rate, which we conjecture to be optimal, improves upon the best known rates previously reported in the literature by Raviv and Karpuk, and generalizes an achievable rate for MDS-TPIR previously found by Freij-Hollanti et al. that is also conjectured to be asymptotically optimal. The setting is then expanded to allow unresponsive  and Byzantine servers. Finally, the scheme is  applied to find a new lower convex hull of (download, upload) pairs of secure and private distributed matrix multiplication that generalizes, and in certain asymptotic settings strictly improves upon the best known previous results.
\end{abstract}

\pagebreak

\section{Introduction}
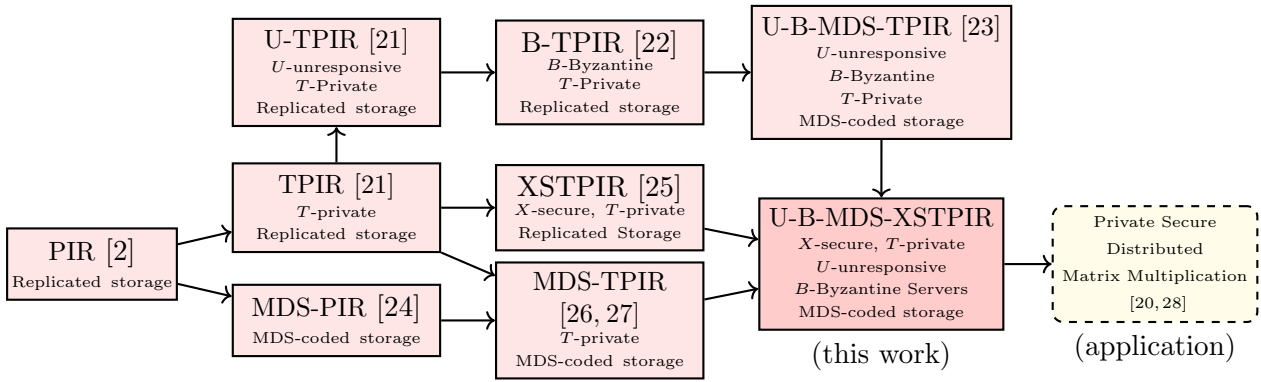
\begin{figure}[h]
\begin{tikzpicture}
\def \theta {5};
\node (PIR) [fill=red!10,draw, thick, shape=rectangle,  anchor=west, text width=2cm] at (0cm,-0.25cm) {\centerline{{\small PIR}\cite{Sun_Jafar_PIR}}\\[-0.3em]\centerline{\tiny Replicated storage}};
\node (TPIR) [fill=red!10, shape=rectangle, draw, thick,   align=center, anchor=west, text width=2.5cm] at (3cm,0.5cm)  {{\small TPIR\cite{Sun_Jafar_TPIR}}\\[2pt] \tiny  $T$-private\\[-5pt] \tiny Replicated storage};

\node (UTPIR) [fill=red!10, shape=rectangle, draw, thick,   align=center, anchor=west, text width=2.5cm] at (3cm,2.3cm)  {{\small U-TPIR\cite{Sun_Jafar_TPIR}}\\[2pt] \tiny  $U$-unresponsive\\[0pt] \tiny $T$-Private\\[-5pt]\tiny Replicated storage};

\node (BTPIR) [fill=red!10, shape=rectangle, draw, thick,  align=center, anchor=west,  text width=2.5cm] at (6.5cm,2.3cm) {B-TPIR \cite{Banawan_Ulukus_BPIR} \\[0pt] \tiny $B$-Byzantine \\[0pt] \tiny $T$-Private\\[-5pt]\tiny Replicated storage};

\node (UBTPIR) [fill=red!10,  shape=rectangle,  align=center, draw, thick, anchor=west] at (9.9cm,2.3cm) {\small U-B-MDS-TPIR \cite{Tajeddine_Gnilke_Karpuk_Hollanti} \\[-5pt] \tiny $U$-unresponsive\\[-5pt] \tiny $B$-Byzantine\\[-5pt]\tiny $T$-Private\\[-5pt]\tiny MDS-coded storage};

\node (MDSPIR) [fill=red!10, shape=rectangle,  anchor=west, draw, thick,   text width=2.5cm] at (3cm,-1cm) {\centerline{{\small MDS-PIR}\cite{Banawan_Ulukus}}\\[-0.3em]\centerline{\tiny MDS-coded storage}};

\node (XSTPIR) [fill=red!10, shape=rectangle, draw, thick,  align=center, anchor=west,  text width=2.5cm] at (6.5cm,0.5cm) {XSTPIR \cite{Jia_Sun_Jafar_XSTPIR} \\[0pt] \tiny $X$-secure, $T$-private \\[-5pt] \tiny Replicated Storage};

\node (MDSTPIR) [fill=red!10, shape=rectangle, draw, thick, align=center, anchor=west, text width=2.5cm] at (6.5cm,-1cm) {\small MDS-TPIR \cite{FREIJ_HOLLANTI, Sun_Jafar_MDSTPIR}\\[1pt] \tiny $T$-private\\[-5pt] \tiny MDS-coded storage};

\node (MDSXSTPIR) [fill=red!20, label=below:(this work), shape=rectangle,  align=center, draw, thick, anchor=west] at (10cm,-0.25cm) {\small U-B-MDS-XSTPIR \\[-5pt] \tiny $X$-secure, $T$-private\\[-5pt] \tiny $U$-unresponsive\\[-5pt] \tiny $B$-Byzantine Servers\\ [-5pt]\tiny MDS-coded storage};
\draw [thick, ->](PIR)--(TPIR);
\draw [thick,->](PIR)--(MDSPIR);
\draw [thick,->](TPIR)--(XSTPIR);
\draw [thick,->](TPIR)--(MDSTPIR);
\draw [thick,->](MDSPIR)--(MDSTPIR);
\draw [thick,->](XSTPIR)--(MDSXSTPIR);
\draw [thick, ->](MDSTPIR)--(MDSXSTPIR);
\draw [thick, ->](TPIR)--(UTPIR);
\draw [thick, ->](UTPIR)--(BTPIR);
\draw [thick, ->](BTPIR)--(UBTPIR);
\draw [thick, ->](UBTPIR)--(MDSXSTPIR);

\node (PSDMM) [right=3cm of MDSXSTPIR, fill=yellow!10, rounded corners, shape=rectangle,  label=below:(application), align=center, draw, dashed, thick, anchor=west] at (10.9cm,-0.25cm) {  \tiny  Private Secure\\[-0.3em]\tiny Distributed\\[-0.3em] \tiny Matrix Multiplication\\[-0.3em]\tiny\cite{Kim_Lee_PSCC, Chang_Tandon_PSDMM}};
\draw [thick, ->](MDSXSTPIR)--(PSDMM);

\end{tikzpicture}
\caption{\small \it The U-B-MDS-XSTPIR setting studied in this work generalizes previously studied settings of PIR \cite{Sun_Jafar_PIR}, TPIR \cite{Sun_Jafar_TPIR}, MDS-PIR \cite{Banawan_Ulukus}, MDS-TPIR \cite{FREIJ_HOLLANTI, Sun_Jafar_MDSTPIR}, XSTPIR \cite{Jia_Sun_Jafar_XSTPIR}, U-TPIR \cite{Sun_Jafar_TPIR}, B-TPIR \cite{Banawan_Ulukus_BPIR}, and  U-B-MDS-TPIR \cite{Tajeddine_Gnilke_Karpuk_Hollanti} as shown, and finds application beyond PIR in the context of Private Secure Distributed Matrix Multiplication (PSDMM).}\label{fig:blocksummary}
\end{figure}

Originating in computer science and cryptography, the problem of private information retrieval (PIR) \cite{PIRfirst} seeks efficient ways to retrieve desired messages from distributed servers without disclosing to the servers which messages are desired. The \emph{rate} of PIR is the maximum number of bits of desired message that can be retrieved per bit of total download from all servers \cite{Sun_Jafar_PIR}. PIR has recently attracted much attention in the information theory community, where the focus has been on finding the capacity (maximum rate) \cite{Sun_Jafar_PIR} or equivalently, minimizing the download cost \cite{Tajeddine_Rouayheb} under various constraints. The study of PIR is important from an information theoretic perspective not only because privacy is important, but also because optimal PIR schemes often reveal novel coding structures, thereby advancing our understanding of structured codes, a cornerstone of network information theory. The fundamental significance of these coding structures is emphasized by the connections between PIR and a number of other important problems such as locally decodable codes \cite{LDC, YekhaninPhd}, locally repairable codes \cite{Gopalan_Huang_Simitci_Yekhanin}, batch codes \cite{Batch}, oblivious transfer \cite{Rabin, SymPIR}, instance hiding \cite{Hide_one, PIRfirst}, secret sharing \cite{Shamir}, blind interference alignment \cite{Jafar_corr, Sun_Jafar_BIAPIR}, and secure computation \cite{yao1982protocols}, including  recent works on secure distributed matrix multiplication \cite{Chang_Tandon_SDMMOS, DOliveira_Rouayheb_Karpuk, Kakar_Ebadifar_Sezgin, Aliasgari_Simeone_Kliewer, Jia_Jafar_SDMM, Chang_Tandon_PSDMM}. As the literature on information theoretic PIR continues to grow, it is also  valuable to find unified perspectives that combine our understanding of various aspects of PIR and allow  generalizations beyond PIR. Against this background, the contribution of this work is summarized  in Figure \ref{fig:blocksummary}.

The capacity of PIR with $K$ messages, $N$ servers, and replicated storage was characterized in \cite{Sun_Jafar_PIR} as $C_{\tiny\mbox{PIR}}=\left(1+\frac{1}{N}+\cdots+\frac{1}{N^{K-1}}\right)^{-1}$. Since the number of messages, $K$ is typically large, of particular interest is the asymptotic value of capacity as $K\rightarrow\infty$. Evidently,  the asymptotic capacity of PIR is $C^\infty_{\tiny\mbox{PIR}}=1-\frac{1}{N}$. The asymptotically optimal achievable scheme builds upon a prior construction from \cite{Shah_Rashmi_Kannan} and may be seen as a form of blind interference alignment \cite{Jafar_corr}.  The capacity of TPIR, i.e., PIR with a $T$-privacy constraint and replicated storage was characterized in \cite{Sun_Jafar_TPIR}.  and its asymptotic value is $C^\infty_{\tiny\mbox{TPIR}}=1-\frac{T}{N}$. The optimal achievable scheme uses an MDS coded query structure. The $T$-Privacy constraint requires that no information about the desired message index is leaked to any set of up to $T$ colluding servers. The capacity of MDS-PIR, i.e., PIR with $(N,K_c)$ MDS-coded storage was characterized in \cite{Banawan_Ulukus} and its asymptotic value turns out to be  $C^\infty_{\tiny\mbox{MDS-PIR}}=1-\frac{K_c}{N}$. MDS-TPIR, i.e., PIR with both $T$-privacy and $(N,K_c)$ MDS coded storage was studied in \cite{FREIJ_HOLLANTI} and while its capacity remains open \cite{Sun_Jafar_MDSTPIR}, the asymptotic achievable rate of $R^\infty_{\tiny\mbox{MDS-TPIR}}=1-\frac{T+K_c-1}{N}$ is expected to be  optimal. The novel achievable scheme of \cite{FREIJ_HOLLANTI} is based on star products of GRS (Generalized Reed-Solomon) codes. The asymptotic capacity of XSTPIR, i.e., PIR with $X$-secure storage, $T$-private queries, and replicated storage was found in  \cite{Jia_Sun_Jafar_XSTPIR} as $1-(X+T)/N$. The achievable scheme of \cite{Jia_Sun_Jafar_XSTPIR} is based on the novel idea of cross-subspace alignment, which has subsequently found use in the context of secure distributed matrix multiplication \cite{Kakar_Ebadifar_Sezgin, Jia_Jafar_SDMM}. 

In this work we study the problem of U-B-MDS-XSTPIR, i.e., PIR with $X$-secure data, $T$-private queries,  $(N,K_c)$ MDS coded storage, where $U$ servers are unresponsive and up to $B$ servers are Byzantine (who may return erroneous responses). In particular we show that a rate of $R^\infty_{\tiny\mbox{U-B-MDS-XSTPIR}}=1-\left(\frac{K_c+X+T+2B-1}{N-U}\right)$ is achievable for any number of messages $K$. This rate strictly improves upon the previous best known rate $R=\left(1-\left(\frac{K_c+X+T+2B-1}{N-U}\right)\right)\left(\frac{K_c}{K_c+X}\right)$ for U-B-MDS-XSTPIR, found\footnote{Reference \cite{Raviv_Karpuk} considers the problem of private polynomial computation with Lagrange encoding, which reduces to U-B-MDS-XSTPIR in the special case where the functions to be computed are all distinct coordinate projections.} in \cite{Raviv_Karpuk}. In fact, for MDS-XSTPIR, i.e., with $U=B=0$, we conjecture that our rate of $R^\infty_{\tiny\mbox{MDS-XSTPIR}}=1-\left(\frac{K_c+X+T-1}{N}\right)$ is  asymptotically optimal as $K\rightarrow\infty$, thus generalizing  a previous conjecture for MDS-TPIR in \cite{FREIJ_HOLLANTI} that can be obtained by further setting $X=0$. Remarkably, U-B-MDS-XSTPIR is a generalization of PIR, TPIR, MDS-PIR, XSTPIR, U-TPIR, B-TPIR, and U-B-MDS-TPIR and the asymptotically optimal (or the best known) structured coding schemes for all of these problems can be obtained as a special case of the unified scheme for U-B-MDS-XSTPIR that we present in this work.  The basis for this unified view, and the central technical contribution of this work, is a  scheme that combines  the cross-subspace alignment idea of \cite{Jia_Sun_Jafar_XSTPIR}  with a layered structure that allows successive decoding and interference cancellation to retrieve multiple layers of  symbols from the desired message.  The scheme is also shown to be applicable to the problem of secure and private distributed matrix multiplication (PSDMM) that was recently introduced in \cite{Kim_Lee_PSCC, Chang_Tandon_PSDMM}. Remarkably,  the new scheme is able to generalize, and in certain asymptotic settings strictly improve upon the previously best known rates for PSDMM.

{\it Notations:} For a positive integer $N$, $[N]$ stands for the set $\{1,2,\dots,N\}$. The notation $X_{[N]}$ denotes the set $\{X_1,X_2,\dots,X_N\}$. For an index set $\mathcal{I}=\{i_1,i_2,\dots,i_n\}$, $X_{\mathcal{I}}$ denotes the set $\{X_{i_1},X_{i_2},\dots,X_{i_n}\}$. For variables $\alpha_n, n\in[N]$ and an arbitrary function $f(\cdot)$, we denote the $N\times 1$ vector whose $n^{th}$ term is $f(\alpha_n)$, as $\vect{f(\alpha)}$.

\section{Problem Statement: U-B-MDS-XSTPIR}
Consider $K$ independent messages, $W_1, W_2,\dots, W_K$. Each message is represented by $\ell$ uniformly random symbols from the finite field $\mathbb{F}_q$.
\begin{align}
H(W_1)=H(W_2)=\dots=H(W_K)=\ell,\label{eq:msgentropy}\\
H(W_{[K]})=K\ell,\label{eq:msgind}
\end{align}
in $q$-ary units. Note that as is typical in information theory, the message sizes are unbounded, and the coding scheme may freely choose the block size $\ell$. The information stored at the $n^{th}$ server is denoted by $S_n$, $n\in[N]$. Messages are stored among $N$ servers according to an MDS$(N,X+K_c)$ code which codes each message separately. From any $X+K_c$ servers, it must be possible to recover all messages.
\begin{align}
H(W_{[K]}|S_{\mathcal{M}})=0, \quad \forall\mathcal{M}\subset[N], |\mathcal{M}|=X+K_c.
\end{align}
The storage requirement at each server is $K\ell/K_c$, i.e.,
\begin{equation}
H(S_n)=\frac{K\ell}{K_c}, \quad \forall n\in[N].
\end{equation}
Thus, compared to replicated storage, the storage requirement is reduced by a factor of $1/K_c$.
$X$-secure storage, $0\leq X\leq N$, guarantees that any $X$ (or fewer) colluding servers learn nothing about the messages.
\begin{align}\label{eq:secur}
I(S_{\mathcal{X}};W_{[K]})=0,\quad \forall\mathcal{X}\subset[N], |\mathcal{X}|=X.
\end{align}

The user privately and uniformly generates the index of his desired message $\theta\in[K]$. To retrieve the desired message privately, the user generates $N$ queries, $Q_{[N]}^{\theta}$. The $n^{th}$ query $Q_{n}^{\theta}$ is sent to the $n^{th}$ server. The user has no prior knowledge of the information stored at the servers, i.e.,
\begin{align}\label{eq:qsind}
I(S_{[N]};\theta,Q_{[N]}^{\theta})=0.
\end{align}
$T$-privacy, $0\leq T\leq N$, guarantees that any $T$ (or fewer) colluding servers learn nothing about the desired message index $\theta$.
\begin{align}\label{eq:priv}
I(Q_{\mathcal{T}}^{\theta},S_{\mathcal{T}};\theta)=0,\quad \forall\mathcal{T}\subset[N], |\mathcal{T}|=T.
\end{align}
Upon receiving the user's query $Q_{n}^{\theta}$, the $n^{th}$ server responds  with the answer $A_n^{\theta}$. 

There exists a set of servers $\mathcal{B}$, $\mathcal{B}\subset[N], |\mathcal{B}|\leq B$, known as Byzantine servers, and another (disjoint) set of servers 
$\mathcal{U}$, $\mathcal{U}\subset[N], |\mathcal{U}|=U$, known as unresponsive servers. The user knows $U, B$ but the realizations of the sets $\mathcal{U},\mathcal{B}$, are  not known to the user \emph{apriori}. The Byzantine servers respond to the user arbitrarily, possibly introducing errors. The unresponsive servers do not respond at all.
However, the remaining servers, i.e., servers in $[N]\setminus(\mathcal{B}\cup\mathcal{U})$,  respond to the user truthfully with a  function of the query and their stored information.
\begin{align}\label{eq:ansfunc}
H(A_n^{\theta}|Q_{n}^{\theta},S_n)=0, \quad \forall n\in[N]\setminus (\mathcal{B}\cup\mathcal{U}).
\end{align}
The user must be able to recover the desired message $W_{\theta}$ from the responses that he receives.
\begin{align}\label{eq:decode}
H(W_{\theta}\mid A_{{[N]\backslash\mathcal{U}}}^{\theta},Q_{[N]}^{\theta},\theta)=0\quad \forall \mathcal{U},\mathcal{B}\subset[N], \mathcal{U}=U, \mathcal{B}=B, \mathcal{U}\cap\mathcal{B}=\emptyset.
\end{align}
The rate of a U-B-MDS-XSTPIR scheme is defined by the number of bits of desired message that are retrieved per total bit of download from all servers on average,
\begin{align}
R_{\text{\tiny U-B-MDS-XSTPIR}}=\frac{H(W_{\theta})}{\sum_{n\in[N]\setminus \mathcal{U}}A_{n}^{\theta}}=\frac{\ell}{D}.
\end{align}
$D=\sum_{n\in[N]\setminus\mathcal{U}}A_{n}^{\theta}$ is the expected number of downloaded bits from all servers. When $B=0, U=0$, i.e., there are no Byzantine servers and no unresponsive servers, then we refer to the problem  simply as MDS-XSTPIR.

\section{Result: An Achievable Rate for U-B-MDS-XSTPIR}
The following lemma is essentially inherited from \cite{Jia_Sun_Jafar_XSTPIR} with minor notational adjustments. Since this lemma is used extensively in this work, a brief proof is also included for the sake of completeness.
\begin{lemma}\label{lemma:csa}
If $\underline{1}, \underline{2}, \cdots, \underline{L}, \alpha_1,\alpha_2,\cdots,\alpha_N$ are $N+L$  distinct elements of $\mathbb{F}_q$, with $1\leq L\leq N-1$, then the following $N\times N$ matrix is invertible over $\mathbb{F}_q$.
\begin{align}
{\bf M}_{L,N}&\triangleq\left[
\begin{matrix}
\frac{1}{\underline{1}-\alpha_1}&\frac{1}{\underline{2}-\alpha_1}&\cdots&\frac{1}{\underline{L}-\alpha_1}&1&\alpha_1&\cdots&\alpha_1^{N-L-1}\\
\frac{1}{\underline{1}-\alpha_2}&\frac{1}{\underline{2}-\alpha_2}&\cdots&\frac{1}{\underline{L}-\alpha_2}&1&\alpha_2&\cdots&\alpha_2^{N-L-1}\\
\cdots&\cdots&\cdots&\cdots&\cdots&\cdots&\cdots&\cdots\\
\frac{1}{\underline{1}-\alpha_N}&\frac{1}{\underline{2}-\alpha_N}&\cdots&\frac{1}{\underline{L}-\alpha_N}&1&\alpha_N&\cdots&\alpha_N^{N-L-1}\\
\end{matrix}
\right]
\end{align}
\end{lemma}
\proof To set up a proof by contradiction, suppose $M$ is not invertible. Then there exist constants $c_n\in\mathbb{F}_q, n\in[N]$, at least one of which is non-zero, such that $\sum_{n\in[N]}c_n{\bf M}_{:,n}={\bf 0}$, where ${\bf M}_{:,n}$ is the $n^{th}$ column of ${\bf M}$.  Define
\begin{align}
\Delta&\triangleq (\underline{1}-\alpha)(\underline{2}-\alpha)\cdots(\underline{L}-\alpha).
\end{align}
Then the polynomial 
\begin{align}
g(\alpha)&=\Delta\left(\sum_{l\in[L]}\frac{c_l}{\underline{l}-\alpha}+\sum_{n=L+1}^{N}c_n\alpha^{n-L-1}\right)
\end{align}
has at least $N$ distinct roots: $\alpha_1, \alpha_2, \cdots, \alpha_N$. But $g(\alpha)$ has degree no more than $N-1$, so it must be the zero polynomial.  This implies that $c_n=0$ for all $n\in[N]$. The contradiction completes the proof. $\hfill\square$

\begin{theorem}\label{thm:lobound}
The following rate is achievable for U-B-MDS-XSTPIR,
\begin{equation}
R_{\text{\tiny U-B-MDS-XSTPIR}}(N,K_c,X,T,U,B,K)= 1-\left(\frac{K_c+X+T+2B-1}{N-U}\right).
\end{equation}
\end{theorem}

 The achievability of this rate,  proved in Section \ref{sec:achi}, is the central contribution of this work. It is based on a coding scheme that uses  cross-subspace alignment along with a layered structure that allows successive decoding with interference cancellation. Note that previously the best known achievable result for U-B-MDS-XSTPIR  for large number of messages ($K\rightarrow\infty$) was $R=\left(1-\left(\frac{K_c+X+T+2B-1}{N-U}\right)\right)\left(\frac{K_c}{K_c+X}\right)$, found in \cite{Raviv_Karpuk}. Evidently our scheme achieves a strictly higher rate. While we conjecture that the rate in Theorem \ref{thm:lobound} for MDS-XSTPIR ($U=0,B=0$) is also the asymptotic capacity of MDS-XSTPIR, a converse proof to this effect remains beyond reach. This is to be expected, because the converse proof has also been unavailable for MDS-TPIR, which is a special case of MDS-XSTPIR.  Our final result appears in Section \ref{sec:PSDMM} where the result of Theorem \ref{thm:lobound} is  applied to the problem of Private Secure Distributed Matrix Multiplication.

\section{Proof of Theorem \ref{thm:lobound}}\label{sec:achi}
First we provide the proof of achievability for $U=0, B=0$, i.e., with no unresponsive or Byzantine servers. Throughout the scheme, let us define
\begin{equation}
L=N-(K_c+X+T-1).
\end{equation}
and let us set
\begin{equation}
\ell=L K_c.
\end{equation}
Let us start with an illustrative example.
\subsection{$X=1, T=1, K_c=2, N=4$}
Here we have $L=1$ and $\ell=2$. So let each message consist of $\ell=2$ symbols from a finite field $\mathbb{F}_q$, where $q\geq L+N=5$. Let $\mathbf{W}_{11}$ and $\mathbf{W}_{12}$ be two $1\times K$ row vectors containing the first and second symbol from every message, respectively. Let $\mathbf{Z}_{11}$ be a uniformly distributed random noise vector from $\mathbb{F}_q^{1\times K}$, that will be used to provide $X=1$ security for the stored data. Let $\mathbf{Z}_{11}^{'1}$, $\mathbf{Z}_{11}^{'2}$ be independent, uniformly distributed random noise vectors from $\mathbb{F}_q^{K\times 1}$ that will be used to provide $T=1$ privacy for the queries. Let $\mathbf{Q}_{\theta}$ be the $\theta$-th column of the $K\times K$ identity matrix, where $\theta$ is the index of desired message. The independence between message, noise vectors, and desired message index $\theta$ is formalized as follows.
\begin{equation}
H(\mathbf{W}_{11}, \mathbf{W}_{12}, \mathbf{Z}_{11}, \mathbf{Z}_{11}^{'1}, \mathbf{Z}_{11}^{'2}, \theta)=H(\mathbf{W}_{11})+H(\mathbf{W}_{12})+H(\mathbf{Z}_{11})+H(\mathbf{Z}_{11}^{'1})+H(\mathbf{Z}_{11}^{'2})+H(\theta).
\end{equation}
Note that by the definition of $\mathbf{W}_{11}$, $\mathbf{W}_{12}$ and $\mathbf{Q}_{\theta}$, the inner products $\mathbf{W}_{11}\mathbf{Q}_{\theta}$ and $\mathbf{W}_{12}\mathbf{Q}_{\theta}$ are precisely the two symbols of the desired message, that the user wishes to retrieve. Let $\underline{1}, \alpha_1, \alpha_2, \cdots, \alpha_N$, represent  $N+1$ distinct elements of $\mathbb{F}_q$. The storage at the $n$-th server is constructed as follows.
\begin{align}
S_n&=\left(\frac{1}{(\underline{1}-\alpha_n)^2}\mathbf{W}_{11}+\frac{1}{\underline{1}-\alpha_n}\mathbf{W}_{12}+\mathbf{Z}_{11}\right),
\end{align}
Thus, the data is coded along with the noise according to an MDS$(N, K_c+X)$, i.e.,  MDS$(4,3)$ code. The presence of noise guarantees that the data is $(X=1)$ secure. The query sent by the user to the $n$-th server to privately retrieve the $\theta$-th message, consists of $K_c=2$ rounds, which are denoted as $Q_n^{\theta,1}$ and $Q_n^{\theta,2}$ respectively.
\begin{align}
Q_n^{\theta,1}=&(\underline{1}-\alpha_n)\mathbf{Q}_{\theta}+(\underline{1}-\alpha_n)^2\mathbf{Z}_{11}^{'1},\\
Q_n^{\theta,2}=&\mathbf{Q}_{\theta}+(\underline{1}-\alpha_n)^2\mathbf{Z}_{11}^{'2}.
\end{align}
Upon receiving the query from user, the answer returned by the $n$-th server is
\begin{equation}
A_n^{\theta}=(S_nQ_n^{\theta,1}, S_nQ_n^{\theta,2}).
\end{equation}
Now let us see why correctness is guaranteed. We rewrite $S_nQ_n^{\theta,1}$ as
\begin{align}
S_nQ_n^{\theta,1}&=\left(\frac{1}{(\underline{1}-\alpha_n)^2}\mathbf{W}_{11}+\frac{1}{\underline{1}-\alpha_n}\mathbf{W}_{12}+\mathbf{Z}_{11}\right)\left((\underline{1}-\alpha_n)\mathbf{Q}_{\theta}+(\underline{1}-\alpha_n)^2\mathbf{Z}_{11}^{'1}\right)\\
&=\frac{1}{\underline{1}-\alpha_n}\mathbf{W}_{11}\mathbf{Q}_{\theta}+\underbrace{\left(\mathbf{W}_{11}\mathbf{Z}_{11}^{'1}+\mathbf{W}_{12}\mathbf{Q}_{\theta}\right)}_{I_1}+(\underline{1}-\alpha_n)\underbrace{\left(\mathbf{W}_{12}\mathbf{Z}_{11}^{'1}+\mathbf{Z}_{11}\mathbf{Q}_{\theta}\right)}_{I_2}+(\underline{1}-\alpha_n)^2\underbrace{\mathbf{Z}_{11}\mathbf{Z}_{11}^{'1}}_{I_3}.
\end{align}
Now, note that the terms $1, (\underline{1}-\alpha_n), (\underline{1}-\alpha_n)^2$, can each be expanded into weighted
sums of the terms $1, \alpha_n, \alpha_n^2$. Re-grouping terms according to this expansion, and collecting $S_nQ_n^{\theta,1}$ terms from the answers received from all $N=4$ servers, we obtain
\begin{align}
\left[\begin{matrix}
S_1Q_1^{\theta,1}\\
S_2Q_2^{\theta,1}\\
S_3Q_3^{\theta,1}\\
S_4Q_4^{\theta,1}
\end{matrix}
\right]&=\left[
\begin{matrix}
\frac{1}{\underline{1}-\alpha_1}&1&\alpha_1&\alpha_1^2\\
\frac{1}{\underline{1}-\alpha_2}&1&\alpha_2&\alpha_2^2\\
\frac{1}{\underline{1}-\alpha_3}&1&\alpha_3&\alpha_3^2\\
\frac{1}{\underline{1}-\alpha_4}&1&\alpha_4&\alpha_4^2
\end{matrix}
\right]\left[
\begin{matrix}
{\bf W}_{11}{\bf Q}_\theta\\
I_1+\underline{1}I_2+\underline{1}^2I_3\\
-I_2-\underline{1}I_3-\underline{1}I_3\\
I_3
\end{matrix}
\right]
\end{align}
Since the $4\times 4$ matrix is ${\bf M}_{1,4}$ which is invertible according to Lemma \ref{lemma:csa}, the user is able to retrieve his first desired symbol, ${\bf W}_{11}{\bf Q}_\theta$. Now, in order to retrieve his second desired symbol, 
  ${\bf W}_{11}{\bf Q}_\theta$, the user will use successive decoding along with  cancellation of interference from the previously retrieved desired symbol. Consider the second part of the answer received from each server, $S_nQ_n^{\theta,2}$, which can be written as follows.
 \begin{align}
S_nQ_n^{\theta,2}&=\left(\frac{1}{(\underline{1}-\alpha_n)^2}\mathbf{W}_{11}+\frac{1}{\underline{1}-\alpha_n}\mathbf{W}_{12}+\mathbf{Z}_{11}\right)\left(\mathbf{Q}_{\theta}+(\underline{1}-\alpha_n)^2\mathbf{Z}_{11}^{'2}\right)\\
&=\frac{1}{(\underline{1}-\alpha_n)^2}\underbrace{\mathbf{W}_{11}\mathbf{Q}_{\theta}}_{I_0'}+\frac{1}{\underline{1}-\alpha_n}\mathbf{W}_{12}\mathbf{Q}_{\theta}\notag\\
&\quad\quad+\underbrace{(\mathbf{W}_{11}\mathbf{Z}_{11}^{'2}+\mathbf{Z}_{11}\mathbf{Q}_{\theta})}_{I_1'}+(\underline{1}-\alpha_n)\underbrace{\mathbf{W}_{12}\mathbf{Z}_{11}^{'2}}_{I_2'}+(\underline{1}-\alpha_n)^2\underbrace{\mathbf{Z}_{11}\mathbf{Z}_{11}^{'2}}_{I_3'}
\end{align}
Aside from the desired symbol ${\bf W}_{12}{\bf Q}_\theta$, there are four \emph{interference} terms $I_0', I_1', I_2', I_3'$. Now, since the user has already retrieved ${\bf W}_{11}{\bf Q}_\theta$, he can subtract $I_0'$ from $S_nQ_n^{\theta,2}$. Furthermore, like before, the remaining interference terms can be expanded along $\alpha_n^t$, $t\in\{0,1,2\}$.
Thus the user is able to obtain
\begin{align}
\left[\begin{matrix}
S_1Q_1^{\theta,2}-\frac{I_0'}{(\underline{1}-\alpha_1)^2}\\
S_2Q_2^{\theta,2}-\frac{I_0'}{(\underline{1}-\alpha_2)^2}\\
S_3Q_3^{\theta,2}-\frac{I_0'}{(\underline{1}-\alpha_3)^2}\\
S_4Q_4^{\theta,2}-\frac{I_0'}{(\underline{1}-\alpha_4)^2}
\end{matrix}
\right]&=\left[
\begin{matrix}
\frac{1}{\underline{1}-\alpha_1}&1&\alpha_1&\alpha_1^2\\
\frac{1}{\underline{1}-\alpha_2}&1&\alpha_2&\alpha_2^2\\
\frac{1}{\underline{1}-\alpha_3}&1&\alpha_3&\alpha_3^2\\
\frac{1}{\underline{1}-\alpha_4}&1&\alpha_4&\alpha_4^2
\end{matrix}
\right]\left[
\begin{matrix}
{\bf W}_{12}{\bf Q}_\theta\\
I_1'+\underline{1}I_2'+\underline{1}^2I_3'\\
-I_2'-\underline{1}I_3'-\underline{1}I_3'\\
I_3'
\end{matrix}
\right]
\end{align}
from which, by inverting the matrix ${\bf M}_{1,4}$, the user is able to retrieve his second desired symbol, ${\bf W}_{12}{\bf Q}_\theta$. This completes the proof of correctness.

\begin{table}[t]
\begin{align}\notag
\begin{array}{cc}\hline\\[-1em]
&\mbox{Server `$n$' (Replace $\alpha$ with $\alpha_n$)} \\[0.1em]\hline\\[0.1em]
\mbox{Storage}(S_n) & \frac{1}{(\underline{1}-\alpha)^2}\mathbf{W}_{11}+\frac{1}{\underline{1}-\alpha}\mathbf{W}_{12}+\mathbf{Z}_{11}\\[+1em]\hline\\[-0.2em]
\mbox{Query} &(\underline{1}-\alpha)\mathbf{Q_\theta}+(\underline{1}-\alpha)^2\mathbf{Z}_{11}^{'1}\\
(Q_n^{[\theta]})&\begin{tikzpicture}\draw[dashed] (0,0)--(5,0);\end{tikzpicture}\\[0.1em]
&\mathbf{Q_\theta}+(\underline{1}-\alpha)^2\mathbf{Z}_{11}^{'2}\\[1em]\hline\\[-0.5em]
\multicolumn{2}{c}{\mbox{Desired symbols appear along vectors}}\\
\multicolumn{2}{c}{\vect{(\underline{1}-\alpha)^{-1}}}\\
\multicolumn{2}{c}{\begin{tikzpicture}\draw[dashed] (0,0)--(5,0);\end{tikzpicture}}\\
\multicolumn{2}{c}{\emphv{\vect{(\underline{1}-\alpha)^{-2}}},\vect{(\underline{1}-\alpha)^{-1}}}\\[-0.5em]\\\hline\\[-0.5em]
\multicolumn{2}{c}{\mbox{Interference  appears along vectors}}\\
\multicolumn{2}{c}{\vect{1}, \vect{(\underline{1}-\alpha)}, \vect{(\underline{1}-\alpha)^{2}}\equiv\vect{1},\vect{\alpha},\vect{\alpha^2}} \\[1em]\hline
\end{array}
\end{align}
\caption{\it\small A summary of the  MDS-XSTPIR scheme for $X=1,T=1, K_c=2, N=4, U=0, B=0$,  showing storage at each server, the queries, and a partitioning of signal and interference dimensions contained in the answers from each server.}\label{table:summary1}
\end{table}
For ease of reference, a compact  summary of the storage at each server, the queries, and a partitioning of signal and interference dimensions contained in the answers from each server, is provided in Table \ref{table:summary1}. Queries and answers of each round are partitioned with dashed  lines. Recovered desired symbols from previous rounds that can be canceled appear along vectors that are wrapped with rounded-corner boxes.

$T=1$-privacy and $X=1$-security follows from the fact that queries and storage are protected by the i.i.d. uniformly distributed noise vectors $\mathbf{Z}_{11}$ and $\mathbf{Z}_{11}^{'1}$, $\mathbf{Z}_{11}^{'2}$ respectively. Finally, let us calculate the rate achieved by the scheme. From $8$ downloaded $q$-ary symbols, the user retrieves $2$ desired $q$-ary symbols, so the rate achieved is $R=2/8=1/4=1-3/4$. This completes the proof of achievability for the setting $U=B=0, X=1, T=1, K_c=2, N=4$.

\subsubsection{$X=1, T=1, K_c=2, N=5$}
Here we have $L=N-(X+T+K_c -1)=2$ and $\ell=LK_c=4$. So let each message consist of $\ell=4$ symbols from a finite field $\mathbb{F}_q$, where $q\geq L+N=7$.  Let $\mathbf{W}_{11}, \mathbf{W}_{21}, \mathbf{W}_{12}, \mathbf{W}_{22}$ be four $1\times K$ row vectors containing the four symbols from every message, respectively. Let $\mathbf{Z}_{11}, \mathbf{Z}_{21}$ be two independent, uniformly distributed random noise vector from $\mathbb{F}_q^{1\times K}$ that will be used to guarantee $X=1$ security. Similarly, let $\mathbf{Z}_{11}^{'1}, \mathbf{Z}_{21}^{'1}$, $\mathbf{Z}_{11}^{'2},\mathbf{Z}_{21}^{'2}$ be independent, uniformly distributed random noise vectors from $\mathbb{F}_q^{K\times 1}$ that will be used to guarantee $T=1$ privacy. As before, let $\mathbf{Q}_{\theta}$ be the $\theta$-th column of the $K\times K$ identity matrix, where $\theta$ is the index of desired message. The desired message $W_{\theta}$ can be represented as
\begin{align}
W_{\theta}&=(\mathbf{W}_{l k}\mathbf{Q}_{\theta})_{l\in[2],k\in[2]}\\
&=({\bf W}_{11}{\bf Q}_\theta, {\bf W}_{12}{\bf Q}_\theta,{\bf W}_{21}{\bf Q}_\theta, {\bf W}_{22}{\bf Q}_\theta).
\end{align}

The independence between message, noise vectors, and desired message index $\theta$ is specified as follows.
\begin{align}
&H(\mathbf{W}_{11},\mathbf{W}_{21},\mathbf{W}_{12}, \mathbf{W}_{22}, \mathbf{Z}_{11}, \mathbf{Z}_{21}, \mathbf{Z}_{11}^{'1},\mathbf{Z}_{21}^{'1},\mathbf{Z}_{11}^{'2}, \mathbf{Z}_{21}^{'2}, \theta)\notag\\
&=\sum_{l\in[2],k\in[2]}H(\mathbf{W}_{l k})+H(\mathbf{Z}_{11})+H(\mathbf{Z}_{21})+H(\mathbf{Z}_{11}^{'1})+H(\mathbf{Z}_{21}^{'1})+H(\mathbf{Z}_{11}^{'2})+H(\mathbf{Z}_{21}^{'2})+H(\theta).
\end{align}
Let $\underline{1},\underline{2},\alpha_1,\alpha_2, \cdots,\alpha_5$ be $L+N=2+5=7$ distinct elements of $\mathbb{F}_q$, $q\geq 7$. The storage at the $n$-th server is constructed as follows.
\begin{equation}
S_n=({S_{n1},S_{n2}}),
\end{equation}
where
\begin{align}
S_{n1}&=\frac{1}{(\underline{1}-\alpha_n)^2}\mathbf{W}_{11}+\frac{1}{\underline{1}-\alpha_n}\mathbf{W}_{12}+\mathbf{Z}_{11},\label{eq:s1}\\
S_{n2}&=\frac{1}{(\underline{2}-\alpha_n)^2}\mathbf{W}_{21}+\frac{1}{\underline{2}-\alpha_n}\mathbf{W}_{22}+\mathbf{Z}_{21}\label{eq:s2}
\end{align}
so that each of \eqref{eq:s1} and \eqref{eq:s2} codes noise with message symbols across $N$ servers according to an MDS$(N, K_c+X)$ code, guaranteeing $X=1$ security on top of MDS coded storage. The query sent to the $n$-th server to retrieve the $\theta^{th}$ message consists of $K_c=2$ rounds, $Q_n^{\theta,1}$ and $Q_n^{\theta,2}$. Furthermore, we will set
\begin{align}
Q_n^{\theta,1}&=(Q_{n1}^{\theta,1},Q_{n2}^{\theta,1})\\
Q_n^{\theta,2}&=(Q_{n1}^{\theta,2},Q_{n2}^{\theta,2})
\end{align}
where
\begin{align}
Q_{n1}^{\theta,1}=&(\underline{1}-\alpha_n)\mathbf{Q}_{\theta}+(\underline{1}-\alpha_n)^2\mathbf{Z}_{11}^{'1},\\
Q_{n2}^{\theta,1}=&(\underline{2}-\alpha_n)\mathbf{Q}_{\theta}+(\underline{2}-\alpha_n)^2\mathbf{Z}_{21}^{'1},\\
Q_{n1}^{\theta,2}=&\mathbf{Q}_{\theta}+(\underline{1}-\alpha_n)^2\mathbf{Z}_{11}^{'2},\\
Q_{n2}^{\theta,2}=&\mathbf{Q}_{\theta}+(\underline{2}-\alpha_n)^2\mathbf{Z}_{21}^{'2}.
\end{align}
Upon receiving the query from user, the answer returned by the $n$-th server is comprised of two symbols,
\begin{align}
A_n^{\theta}&=(A_{n1}^\theta, A_{n2}^\theta)\\
&=(S_{n1}Q_{n1}^{\theta,1}+S_{n2}Q_{n2}^{\theta,1}, ~~S_{n1}Q_{n1}^{\theta,2}+S_{n2}Q_{n2}^{\theta,2}).
\end{align}
Now let us see why correctness is guaranteed. Consider the first symbol, $A_{n1}^{\theta}$.
\begin{align}
A_{n1}^\theta
&=S_{n1}Q_{n1}^{\theta,1}+S_{n2}Q_{n2}^{\theta,1}\notag\\
&=\frac{1}{\underline{1}-\alpha_n}\mathbf{W}_{11}\mathbf{Q}_{\theta}+\frac{1}{\underline{2}-\alpha_n}\mathbf{W}_{21}\mathbf{Q}_{\theta}+(\mathbf{W}_{11}\mathbf{Z}_{11}^{'1}+\mathbf{W}_{21}\mathbf{Z}_{21}^{'1}+\mathbf{W}_{12}\mathbf{Q}_{\theta}+\mathbf{W}_{22}\mathbf{Q}_{\theta})\notag\\
&\quad\quad+(\underline{1}-\alpha_n)(\mathbf{W}_{12}\mathbf{Z}_{11}^{'1}+\mathbf{Z}_{11}\mathbf{Q}_{\theta})+(\underline{2}-\alpha_n)(\mathbf{W}_{22}\mathbf{Z}_{21}^{'1}+\mathbf{Z}_{21}\mathbf{Q}_{\theta})\notag\\
&\quad\quad+(\underline{1}-\alpha_n)^2\mathbf{Z}_{11}\mathbf{Z}_{11}^{'1}+(\underline{1}-\alpha_n)^2\mathbf{Z}_{21}\mathbf{Z}_{21}^{'1}.\label{eq:expandan1}
\end{align}
The first two terms in \eqref{eq:expandan1} are desired message symbols. Each of the remaining $5$ terms can be expanded into weighted sums of terms of the form $\alpha_n^t$, $t\in\{0, 1, 2\}$, allowing the user to represent the symbols $A_{n1}^{\theta}$ downloaded from all $n\in[N]$ servers, as
\begin{align}
\left[\begin{matrix}
A_{11}^{\theta}\\
A_{21}^{\theta}\\
A_{31}^{\theta}\\
A_{41}^{\theta}\\
A_{51}^{\theta}
\end{matrix}
\right]&=\begin{bmatrix}
\frac{1}{\underline{1}-\alpha_1}&\frac{1}{\underline{2}-\alpha_1}&1&\alpha_1&\alpha_1^2\\
\frac{1}{\underline{1}-\alpha_2}&\frac{1}{\underline{2}-\alpha_2}&1&\alpha_2&\alpha_2^2\\
\frac{1}{\underline{1}-\alpha_3}&\frac{1}{\underline{2}-\alpha_3}&1&\alpha_3&\alpha_3^2\\
\frac{1}{\underline{1}-\alpha_4}&\frac{1}{\underline{2}-\alpha_4}&1&\alpha_4&\alpha_4^2\\
\frac{1}{\underline{1}-\alpha_5}&\frac{1}{\underline{2}-\alpha_5}&1&\alpha_5&\alpha_5^2
\end{bmatrix}
\left[
\begin{matrix}
{\bf W}_{11}{\bf Q}_\theta\\
{\bf W}_{21}{\bf Q}_\theta\\
*\\
*\\
*
\end{matrix}
\right]\label{eq:M25}
\end{align}
where we have used $*$ to represent various combinations of interference symbols that can be found explicitly by expanding \eqref{eq:expandan1}, since those forms are not important. What matters is that the $5\times 5$ square matrix  in \eqref{eq:M25} is ${\bf M}_{2,5}$ which is invertible according to Lemma \ref{lemma:csa}, so the user can retrieve the two desired symbols, ${\bf W}_{11}{\bf Q}_\theta$, ${\bf W}_{21}{\bf Q}_\theta$ by inverting the matrix. Next, the user needs to retrieve the remaining two desired symbols ${\bf W}_{12}{\bf Q}_\theta$, ${\bf W}_{22}{\bf Q}_\theta$, for which we will use successive decoding with interference cancellation.
Consider the downloaded symbol  $A_{n2}^{\theta}$.
\begin{align}
A_{n2}^\theta&=S_{n1}Q_{n1}^{\theta,2}+S_{n2}Q_{n2}^{\theta,2}\notag\\
&=\frac{1}{(\underline{1}-\alpha_n)^2}\mathbf{W}_{11}\mathbf{Q}_{\theta}+\frac{1}{(\underline{2}-\alpha_n)^2}\mathbf{W}_{21}{\bf Q}_\theta+\frac{1}{\underline{1}-\alpha_n}\mathbf{W}_{12}\mathbf{Q}_{\theta}+\frac{1}{\underline{2}-\alpha_n}\mathbf{W}_{22}\mathbf{Q}_{\theta}\notag\\
&\quad\quad+(\mathbf{W}_{11}\mathbf{Z}_{11}^{'2}+\mathbf{W}_{21}\mathbf{Z}_{21}^{'2}+\mathbf{Z}_{11}\mathbf{Q}_{\theta}+\mathbf{Z}_{21}\mathbf{Q}_{\theta})+(\underline{1}-\alpha_n)\mathbf{W}_{12}\mathbf{Z}_{11}^{'2}+(\underline{2}-\alpha_n)\mathbf{W}_{22}\mathbf{Z}_{21}^{'2}\notag\\
&\quad\quad+(\underline{1}-\alpha_n)^2\mathbf{Z}_{11}\mathbf{Z}_{11}^{'2}+(\underline{2}-\alpha_n)^2\mathbf{Z}_{21}\mathbf{Z}_{21}^{'2}.\label{eq:expandan2}
\end{align}
The first two symbols in \eqref{eq:expandan2} are desired symbols that have already been decoded. So these terms can be subtracted out, leaving the user with the following downloaded information from all $N=5$ servers.
\begin{align}
\left[\begin{matrix}
A_{12}^{\theta}-\frac{1}{(\underline{1}-\alpha_1)^2}\mathbf{W}_{11}\mathbf{Q}_{\theta}-\frac{1}{(\underline{2}-\alpha_1)^2}\mathbf{W}_{21}{\bf Q}_\theta\\
A_{22}^{\theta}-\frac{1}{(\underline{1}-\alpha_2)^2}\mathbf{W}_{11}\mathbf{Q}_{\theta}-\frac{1}{(\underline{2}-\alpha_2)^2}\mathbf{W}_{21}{\bf Q}_\theta\\
A_{32}^{\theta}-\frac{1}{(\underline{1}-\alpha_3)^2}\mathbf{W}_{11}\mathbf{Q}_{\theta}-\frac{1}{(\underline{2}-\alpha_3)^2}\mathbf{W}_{21}{\bf Q}_\theta\\
A_{42}^{\theta}-\frac{1}{(\underline{1}-\alpha_4)^2}\mathbf{W}_{11}\mathbf{Q}_{\theta}-\frac{1}{(\underline{2}-\alpha_4)^2}\mathbf{W}_{21}{\bf Q}_\theta\\
A_{52}^{\theta}-\frac{1}{(\underline{1}-\alpha_5)^2}\mathbf{W}_{11}\mathbf{Q}_{\theta}-\frac{1}{(\underline{2}-\alpha_5)^2}\mathbf{W}_{21}{\bf Q}_\theta
\end{matrix}
\right]&=\begin{bmatrix}
\frac{1}{\underline{1}-\alpha_1}&\frac{1}{\underline{2}-\alpha_1}&1&\alpha_1&\alpha_1^2\\
\frac{1}{\underline{1}-\alpha_2}&\frac{1}{\underline{2}-\alpha_2}&1&\alpha_2&\alpha_2^2\\
\frac{1}{\underline{1}-\alpha_3}&\frac{1}{\underline{2}-\alpha_3}&1&\alpha_3&\alpha_3^2\\
\frac{1}{\underline{1}-\alpha_4}&\frac{1}{\underline{2}-\alpha_4}&1&\alpha_4&\alpha_4^2\\
\frac{1}{\underline{1}-\alpha_5}&\frac{1}{\underline{2}-\alpha_5}&1&\alpha_5&\alpha_5^2
\end{bmatrix}
\left[
\begin{matrix}
{\bf W}_{12}{\bf Q}_\theta\\
{\bf W}_{22}{\bf Q}_\theta\\
*\\
*\\
*
\end{matrix}
\right]\label{eq:M25b}
\end{align}
Once again, the $5\times 5$ square matrix  in \eqref{eq:M25b} is ${\bf M}_{2,5}$ which is invertible according to Lemma \ref{lemma:csa}, so the user can retrieve his remaining two desired symbols, ${\bf W}_{12}{\bf Q}_\theta$, ${\bf W}_{22}{\bf Q}_\theta$ by inverting the matrix. This completes the proof of correctness. Let us summarize  the storage at each server, the queries, and the partitioning of signal and interference dimensions contained in the answers from each server in Table \ref{table:summary2}.
\begin{table}[!b]
\begin{align}\notag
\begin{array}{cc}\hline
&\mbox{Server `$n$' (Replace $\alpha$ with $\alpha_n$)} \\\hline
\mbox{Storage} & \frac{1}{(\underline{1}-\alpha)^2}\mathbf{W}_{11}+\frac{1}{\underline{1}-\alpha}\mathbf{W}_{12}+\mathbf{Z}_{11}\\
(S_n)&\frac{1}{(\underline{2}-\alpha)^2}\mathbf{W}_{21}+\frac{1}{\underline{2}-\alpha}\mathbf{W}_{22}+\mathbf{Z}_{21}\\\hline
\mbox{Query}&(\underline{1}-\alpha)\mathbf{Q_\theta}+(\underline{1}-\alpha)^2\mathbf{Z}_{11}^{'1}\\
(Q_n^{[\theta]})&(\underline{2}-\alpha)\mathbf{Q_\theta}+(\underline{2}-\alpha)^2\mathbf{Z}_{21}^{'1}\\
&\begin{tikzpicture}\draw[dashed] (0,0)--(5,0);\end{tikzpicture}\\
&\mathbf{Q_\theta}+(\underline{1}-\alpha)^2\mathbf{Z}_{11}^{'2}\\
&\mathbf{Q_\theta}+(\underline{2}-\alpha)^2\mathbf{Z}_{21}^{'2}\\\hline
\multicolumn{2}{c}{\mbox{Desired symbols appear along vectors}}\\
\multicolumn{2}{c}{\vect{(\underline{1}-\alpha)^{-1}},\vect{(\underline{2}-\alpha)^{-1}}}\\
\multicolumn{2}{c}{\mbox{---------------------------------------------------------------------------------}}\\
\multicolumn{2}{c}{\emphv{\vect{(\underline{1}-\alpha)^{-2}}},\emphv{\vect{(\underline{2}-\alpha)^{-2}}},\vect{(\underline{1}-\alpha)^{-1}},\vect{(\underline{2}-\alpha)^{-1}}}\\
\multicolumn{2}{c}{\mbox{Interference  appears along vectors}}\\
\multicolumn{2}{c}{\vect{1}, \vect{(\underline{1}-\alpha)}, \vect{(\underline{1}-\alpha)^{2}},\vect{(\underline{2}-\alpha)},\vect{(\underline{2}-\alpha)^{2}}\equiv\vect{1},\vect{\alpha},\vect{\alpha^2}} \\\hline
\end{array}
\end{align}
\caption{\it\small A summary of the  MDS-XSTPIR scheme for $X=1,T=1, K_c=2, N=5, U=0, B=0$,  showing storage at each server, the queries, and a partitioning of signal and interference dimensions contained in the answers from each server.}\label{table:summary2}
\end{table}
$T=1$-privacy and $X=1$-security follows from the fact that queries and storage are protected by the i.i.d. uniformly distributed noise vectors. Now consider the rate achieved by the scheme. Since the user downloads $2$ symbols from each of $5$ servers, we note that from a total of $10$ downloaded $q$-ary symbols, the user is able to recover $4$ desired $q$-ary symbols, so the rate achieved is $R=4/10=2/5=1-3/5$. This completes the construction of the scheme for the setting $U=B=0, X=1, T=1, K_c=2, N=5$. We now specify the scheme for  $U=B=0$ and arbitrary $X, T, K_c, N$ parameters.

\subsubsection{$U=B=0$, arbitrary $X, T, K_c, N$}
Let each message consist of $\ell=LK_c$ symbols from a finite field $\mathbb{F}_q$ where  $L=N-(X+T+K_c-1)$ and $q\geq L+N$. Let $\mathbf{W}_{l k}, l\in[L], k\in[K_c]$ be $1\times K$ row vectors. For each value of $\l\in[L], k\in[K_c]$, the $1\times K$ row vector $\mathbf{W}_{l k}$ contains the $(L(k-1)+l)^{th}$ symbol from every message. Let $(\mathbf{Z}_{l x})_{l\in[L], x\in[X]}$ be independent, uniformly distributed random noise vectors from $\mathbb{F}_q^{1\times K}$ that will be used to guarantee $X$-security. Let $(\mathbf{Z}_{l t}^{'\kappa})_{l\in[L], t\in[T], \kappa\in[K_c]}$ be independent, uniformly distributed random noise vectors from $\mathbb{F}_q^{K\times 1}$ that will be used to guarantee that the queries are $T$-private. As before, let $\mathbf{Q}_{\theta}$ be the $\theta$-th column of the $K\times K$ identity matrix, where $\theta$ is the index of desired message. The desired message $W_{\theta}$ can be represented as,
\begin{align}
W_{\theta}&=({\bf W}_{lk}{\bf Q}_\theta)_{l\in[L], k\in[K_c]}\\
&=\left(\begin{matrix}\mathbf{W}_{11}\mathbf{Q}_{\theta},& \mathbf{W}_{12}\mathbf{Q}_{\theta},& \cdots, &\mathbf{W}_{1K_c}\mathbf{Q}_{\theta}\\ \mathbf{W}_{21}{\bf Q}_\theta,&\mathbf{W}_{22}{\bf Q}_\theta,& \cdots,& \mathbf{W}_{2K_c}{\bf Q}_\theta\\ 
\cdots&\cdots&\cdots&\cdots\\
\mathbf{W}_{L1}\mathbf{Q}_{\theta},& \mathbf{W}_{L2}\mathbf{Q}_{\theta},&\cdots,&\mathbf{W}_{LK_c}\mathbf{Q}_{\theta}
\end{matrix}
\right).
\end{align}
The independence between messages, noise vectors and $\theta$ is formalized as follows.
\begin{align}
&H((\mathbf{W}_{l k})_{l\in[L],k\in[K_c]}, (\mathbf{Z}_{l x})_{l\in[L], x\in[X]}, (\mathbf{Z}_{l t}^{'\kappa})_{l\in[L], t\in[T], \kappa\in[K_c]}, \theta)\notag\\
&=\sum_{l\in[L], k\in[K_c]}H(\mathbf{W}_{l k})+\sum_{l\in[L], x\in[X]}H(\mathbf{Z}_{l x})+\sum_{l\in[L], t\in[T], \kappa\in[K_c]}H(\mathbf{Z}_{l t}^{'\kappa})+H(\theta).
\end{align}
Let $\underline{1}, \underline{2}, \cdots,\underline{L}, \alpha_1,\alpha_2, \cdots,\alpha_N$ be $L+N$ distinct elements of $\mathbb{F}_q$. Since $q\geq N+L$, these constants must exist. The storage at the $n^{th}$ server is comprised of $L$ symbols $(S_{nl})_{l\in[L]}$, i.e.,
\begin{equation}
S_n=(S_{n1},S_{n,2},\dots,S_{nL}).
\end{equation}
For all $l\in[L]$, $S_{nl}$ is constructed as
\begin{align}
S_{nl}&=\frac{1}{(\underline{l}-\alpha_n)^{K_c}}\mathbf{W}_{l 1}+\frac{1}{(\underline{l}-\alpha_n)^{K_c-1}}\mathbf{W}_{l 2}+\dots+\frac{1}{\underline{l}-\alpha_n}\mathbf{W}_{l K_c}+\sum_{x\in[X]}(\underline{l}-\alpha_n)^{x-1}\mathbf{Z}_{l x}\\
&=\sum_{k\in[K_c]}\frac{1}{(\underline{l}-\alpha_n)^{K_c-k+1}}\mathbf{W}_{l k}+\sum_{x\in[X]}(\underline{l}-\alpha_n)^{x-1}\mathbf{Z}_{l x}.\label{eq:schemestor}
\end{align}
Thus, for each $l\in[L]$, the values $S_{nl}$ stored across all $N$ servers comprise an MDS$(N, K_c+X)$ code which includes $X$ noise symbols for $X$-security. The query sent by the user to the $n$-th server, in order to retrieve the $\theta^{th}$ desired message, is comprised of $K_c$ rounds, $(Q_n^{\theta, \kappa})_{\kappa\in[K_c]}$. For each $\kappa\in[K_c]$, the query is constructed as follows.
\begin{equation}
Q_n^{\theta,\kappa}=(Q_{n1}^{\theta,\kappa},Q_{n2}^{\theta,\kappa},\dots,Q_{nL}^{\theta,\kappa}),
\end{equation}
where $\forall l\in[L]$, let us set
\begin{equation}
Q_{nl}^{\theta,\kappa}=(\underline{l}-\alpha_n)^{K_c-\kappa}\mathbf{Q}_{\theta}+\sum_{t\in[T]}(\underline{l}-\alpha_n)^{K_c+t-1}\mathbf{Z}_{l t}^{'\kappa}.
\end{equation}
Upon receiving the query from the user, the $n$-th server responds with the following $K_c$ symbols.
\begin{align}
A_n^{\theta}&=(A_{n1}^\theta, A_{n2}^\theta,\cdots,A_{nK_c}^\theta)
\end{align}
where for all $\kappa\in[K_c]$,
\begin{align}
A_{n\kappa}^\theta&=(S_{n1}Q_{n1}^{\theta,\kappa}+S_{n2}Q_{n2}^{\theta,
\kappa}+\cdots+S_{nL}Q_{nL}^{\theta,\kappa}).
\end{align}
To show that the scheme is correct, for any $\kappa\in[K_c]$, let us rewrite the symbol $A_{n\kappa}^\theta$ as,
\begin{align}
A_{n\kappa}^\theta&=\sum_{l\in[L]}S_{nl}Q_{nl}^{\theta,\kappa}\\
&=\sum_{l\in[L]}\left(\sum_{k\in[K_c]}\frac{1}{(\underline{l}-\alpha_n)^{K_c-k+1}}\mathbf{W}_{l k}+\sum_{x\in[X]}(\underline{l}-\alpha_n)^{x-1}\mathbf{Z}_{l x}\right)\notag\\
&\quad\quad\quad\quad\quad\left((\underline{l}-\alpha_n)^{K_c-\kappa}\mathbf{Q}_{\theta}+\sum_{t\in[T]}(\underline{l}-\alpha_n)^{K_c+t-1}\mathbf{Z}_{l t}^{'\kappa}\right)\\
&=\sum_{l\in[L]}\sum_{k\in[\kappa]}\frac{1}{(\underline{l}-\alpha_n)^{\kappa-k+1}}\mathbf{W}_{l k}\mathbf{Q}_{\theta}+\sum_{l\in[L]}\sum_{k=\kappa+1}^{K_c}(\underline{l}-\alpha_n)^{k-\kappa-1}\mathbf{W}_{l k}\mathbf{Q}_{\theta}\notag\\
&\quad+\sum_{l\in[L]}\sum_{x\in[X]}(\underline{l}-\alpha_n)^{K_c-\kappa+x-1}\mathbf{Z}_{l x}\mathbf{Q}_{\theta}+\sum_{l\in[L]}\sum_{k\in[K_c]}\sum_{t\in[T]}(\underline{l}-\alpha_n)^{k+t-2}\mathbf{W}_{l k}\mathbf{Z}_{l t}^{'\kappa}\notag\\
&\quad+\sum_{l\in[L]}\sum_{x\in[X]}\sum_{t\in[T]}(\underline{l}-\alpha_n)^{K_c+t+x-2}\mathbf{Z}_{l x}\mathbf{Z}_{l t}^{'\kappa}.
\end{align}
Now we will see why it is possible to recover all desired symbols $(\mathbf{W}_{l k}\mathbf{Q}_{\theta})_{l\in[L], k\in[K_c]}$. Consider $\kappa=1$.
\begin{align}
A_{n1}^\theta&=\sum_{l\in[L]}S_{nl}Q_{nl}^{\theta,1}\\
&=\sum_{l\in[L]}\frac{1}{\underline{l}-\alpha_n}\mathbf{W}_{l 1}\mathbf{Q}_{\theta}+\sum_{l\in[L]}\sum_{k=2}^{K_c}(\underline{l}-\alpha_n)^{k-2}\mathbf{W}_{lk}\mathbf{Q}_{\theta}+\sum_{l\in[L]}\sum_{x\in[X]}(\underline{l}-\alpha_n)^{K_c+x-2}\mathbf{Z}_{l x}\mathbf{Q}_{\theta}\notag\\
&\quad+\sum_{l\in[L]}\sum_{k\in[K_c]}\sum_{t\in[T]}(\underline{l}-\alpha_n)^{t+k-2}\mathbf{W}_{lk}\mathbf{Z}_{l t}^{'1}+\sum_{l\in[L]}\sum_{x\in[X]}\sum_{t\in[T]}(\underline{l}-\alpha_n)^{K_c+t+x-2}\mathbf{Z}_{l x}\mathbf{Z}_{l t}^{'1}
\end{align}
The first term contains  the $L$ desired symbols $(\mathbf{W}_{11}\mathbf{Q}_{\theta},\dots,\mathbf{W}_{L1}\mathbf{Q}_{\theta})$ that are to be retrieved in the first round, i.e., for $\kappa=1$.
Each of the remaining four terms constitute interference which can be expanded into weighted sums of terms of the form $\alpha_n^t$, $t\in\{0,1,\dots,K_c+X+T-2\}$. Therefore, collecting the $A_{n1}^\theta$ symbols from all $N$ servers, the user obtains
\begin{align}
\left[
\begin{matrix}
A_{11}^\theta\\
A_{21}^\theta\\
\vdots\\
A_{N1}^\theta
\end{matrix}
\right]&=\begin{bmatrix}
\frac{1}{\underline{1}-\alpha_1}&\cdots&\frac{1}{\underline{L}-\alpha_1}&1&\alpha_1&\cdots&\alpha_1^{K_c+X+T-2}\\
\frac{1}{\underline{1}-\alpha_2}&\cdots&\frac{1}{\underline{L}-\alpha_2}&1&\alpha_2&\cdots&\alpha_2^{K_c+X+T-2}\\
\vdots&\vdots&\vdots&\vdots&\vdots&\vdots&\vdots&\\
\frac{1}{\underline{1}-\alpha_N}&\cdots&\frac{1}{\underline{L}-\alpha_N}&1&\alpha_N&\cdots&\alpha_N^{K_c+X+T-2}\\
\end{bmatrix}
\begin{bmatrix}
{\bf W}_{11}{\bf Q}_\theta\\
{\bf W}_{21}{\bf Q}_\theta\\
\vdots\\
{\bf W}_{L1}{\bf Q}_\theta\\
*\\
\vdots\\
*
\end{bmatrix}\label{eq:MLN}
\end{align}
where $*$ represents various combinations of interference terms, whose precise forms are inconsequential. What matters is that the $N\times N$ matrix in \eqref{eq:MLN} is ${\bf M}_{L,N}$ which is invertible according to Lemma \ref{lemma:csa}, so that the user is able to retrieve the desired symbols $(\mathbf{W}_{11}\mathbf{Q}_{\theta},\dots,\mathbf{W}_{L1}\mathbf{Q}_{\theta})$ by inverting the matrix. 

The scheme proceeds similarly to retrieves desired symbols $(\mathbf{W}_{1\kappa}\mathbf{Q}_{\theta},\dots,\mathbf{W}_{L\kappa}\mathbf{Q}_{\theta})$ with the $\kappa^{th}$ round of queries. To prove this by induction, let us consider any $\kappa$, such that  $2\leq \kappa\leq K_c$, and assume that the desired symbols $(\mathbf{W}_{lk}\mathbf{Q}_{\theta})_{l\in[L], k\in[\kappa-1]}$ have already been retrieved. Now we wish to show that   the desired symbols $(\mathbf{W}_{l\kappa}\mathbf{Q}_{\theta})_{l\in[L]}$ can be retrieved.
\begin{align}
A_{n\kappa}^\theta&=\sum_{l\in[L]}S_{nl}Q_{nl}^{\theta,\kappa}\\
&=\sum_{l\in[L]}\sum_{k\in[\kappa-1]}\frac{1}{(\underline{l}-\alpha_n)^{\kappa-k+1}}\mathbf{W}_{l k}\mathbf{Q}_{\theta}+\sum_{l\in[L]}\frac{1}{\underline{l}-\alpha_n}\mathbf{W}_{l\kappa}\mathbf{Q}_{\theta}\notag\\
&\quad+\sum_{l\in[L]}\sum_{k=\kappa+1}^{K_c}(\underline{l}-\alpha_n)^{k-\kappa-1}\mathbf{W}_{l k}\mathbf{Q}_{\theta}+\sum_{l\in[L]}\sum_{x\in[X]}(\underline{l}-\alpha_n)^{K_c-\kappa+x-1}\mathbf{Z}_{l x}\mathbf{Q}_{\theta}\\
&\quad+\sum_{l\in[L]}\sum_{k\in[K_c]}\sum_{t\in[T]}(\underline{l}-\alpha_n)^{t+k-2}\mathbf{W}_{l k}\mathbf{Z}_{l t}^{'\kappa}+\sum_{l\in[L]}\sum_{x\in[X]}\sum_{t\in[T]}(\underline{l}-\alpha_n)^{K_c+t+x-2}\mathbf{Z}_{l x}\mathbf{Z}_{l t}^{'\kappa}.\label{eq:achelm}
\end{align}
The first term contains symbols that have already been retrieved, so the user can subtract this term from $A_{n\kappa}^\theta$. 
\begin{align}
A_{n\kappa}^{\theta'}&=A_{n\kappa}^\theta-\sum_{l\in[L]}\sum_{k\in[\kappa-1]}\frac{1}{(\underline{l}-\alpha_n)^{\kappa-k+1}}\mathbf{W}_{l k}\mathbf{Q}_{\theta}.
\end{align}
The next term is comprised  of the $L$ symbols $(\mathbf{W}_{l\kappa}\mathbf{Q}_{\theta})_{l\in[L]}$ that the user wishes to retrieve. The remaining $4$ terms constitute interference which can be expanded as before into weighted sums of terms of the form $\alpha_n^t, t\in\{0,1,\ldots, K_c+X+T-2\}$. Therefore, collecting  the $A_{n\kappa}^{\theta'}$ symbols from all $N$ servers, the user obtains,
\begin{align}
\left[
\begin{matrix}
A_{11}^{\theta'}\\
A_{21}^{\theta'}\\
\vdots\\
A_{N1}^{\theta'}
\end{matrix}
\right]&=\begin{bmatrix}
\frac{1}{\underline{1}-\alpha_1}&\cdots&\frac{1}{\underline{L}-\alpha_1}&1&\alpha_1&\cdots&\alpha_1^{K_c+X+T-2}\\
\frac{1}{\underline{1}-\alpha_2}&\cdots&\frac{1}{\underline{L}-\alpha_2}&1&\alpha_2&\cdots&\alpha_2^{K_c+X+T-2}\\
\vdots&\vdots&\vdots&\vdots&\vdots&\vdots&\vdots&\\
\frac{1}{\underline{1}-\alpha_N}&\cdots&\frac{1}{\underline{L}-\alpha_N}&1&\alpha_N&\cdots&\alpha_N^{K_c+X+T-2}\\
\end{bmatrix}
\begin{bmatrix}
{\bf W}_{1\kappa}{\bf Q}_\theta\\
{\bf W}_{2\kappa}{\bf Q}_\theta\\
\vdots\\
{\bf W}_{L\kappa}{\bf Q}_\theta\\
*\\
\vdots\\
*
\end{bmatrix}\label{eq:MLNgen}
\end{align}

The desired symbols $(\mathbf{W}_{l\kappa}\mathbf{Q}_{\theta})_{l\in[L]}$ can be retrieved by inverting the $N\times N$ square matrix in \eqref{eq:MLNgen}, which is guaranteed to be invertible according to Lemma \ref{lemma:csa}. Thus, the induction argument shows that all $\ell=LK_c$ desired symbols are retrieved successfully. A summary of the storage at each server, the queries, and a partitioning of signal and interference dimensions contained in the answers from each server is provided in  Table \ref{table:summarygen}.

$T$-privacy is guaranteed because $\mathbf{Q}_{\theta}$ is protected by the noise vectors $(\mathbf{Z}_{l t}^{'\kappa})_{l\in[L], t\in[T], \kappa\in[K_c]}$ that are i.i.d. uniform and coded according to an MDS$(N,T)$ code. Similarly, $X$-security is guaranteed because for each $l\in[L]$, the messages $(\mathbf{W}_{lk})_{k\in[K_c]}$ are protected by the noise vectors $(\mathbf{Z}_{l x})_{x\in[X]}$ that are i.i.d. uniform and coded according to an MDS$(N,X)$ code. Now let us consider the rate achieved by the scheme. From a total of $NK_c$ downloaded $q$-ary symbols, the user is able to retrieve his $\ell=LK_c$ desired symbols, so the rate achieved is
\begin{equation}
R=\frac{LK_c}{NK_c}=\frac{L}{N}=1-\left(\frac{K_c+X+T-1}{N}\right),
\end{equation}
which matches the result in Theorem \ref{thm:lobound}.

\begin{table}[H]
\begin{align}\notag
\begin{array}{cc}\hline
&\mbox{Server `$n$' (Replace $\alpha$ with $\alpha_n$)} \\\hline
\mbox{Storage} & \frac{1}{(\underline{1}-\alpha)^{K_c}}\mathbf{W}_{11}+\cdots+\frac{1}{\underline{1}-\alpha}\mathbf{W}_{1K_c}+\mathbf{Z}_{11}+\cdots+(\underline{1}-\alpha)^{X-1}\mathbf{Z}_{1X}\\
(S_n)&\frac{1}{(\underline{2}-\alpha)^{K_c}}\mathbf{W}_{21}+\cdots+\frac{1}{\underline{2}-\alpha}\mathbf{W}_{2K_c}+\mathbf{Z}_{21}+\cdots+(\underline{2}-\alpha)^{X-1}\mathbf{Z}_{2X}\\
&\vdots\\
&\frac{1}{(\underline{L}-\alpha)^{K_c}}\mathbf{W}_{L1}+\cdots+\frac{1}{\underline{L}-\alpha}\mathbf{W}_{L'K_c}+\mathbf{Z}_{L1}+\cdots+(\underline{L}-\alpha)^{X-1}\mathbf{Z}_{LX}\\\hline
\mbox{Query}&(\underline{1}-\alpha)^{K_c-1}\mathbf{Q_\theta}+(\underline{1}-\alpha)^{K_c}\mathbf{Z}_{11}^{'1}+\cdots+(\underline{1}-\alpha)^{K_c+T-1}\mathbf{Z}_{1T}^{'1}\\
(Q_n^{[\theta]})&(\underline{2}-\alpha)^{K_c-1}\mathbf{Q_\theta}+(\underline{2}-\alpha)^{K_c}\mathbf{Z}_{21}^{'1}+\cdots+(\underline{2}-\alpha)^{K_c+T-1}\mathbf{Z}_{2T}^{'1}\\
&\vdots\\
&(\underline{L}-\alpha)^{K_c-1}\mathbf{Q_\theta}+(\underline{L}-\alpha)^{K_c}\mathbf{Z}_{L1}^{'1}+\cdots+(\underline{L}-\alpha)^{K_c+T-1}\mathbf{Z}_{LT}^{'1}\\
&\begin{tikzpicture}\draw[dashed] (0,0)--(9,0);\end{tikzpicture}\\
&(\underline{1}-\alpha)^{K_c-2}\mathbf{Q_\theta}+(\underline{1}-\alpha)^{K_c}\mathbf{Z}_{11}^{'2}+\cdots+(\underline{1}-\alpha)^{K_c+T-1}\mathbf{Z}_{1T}^{'2}\\
&(\underline{2}-\alpha)^{K_c-2}\mathbf{Q_\theta}+(\underline{2}-\alpha)^{K_c}\mathbf{Z}_{21}^{'2}+\cdots+(\underline{2}-\alpha)^{K_c+T-1}\mathbf{Z}_{2T}^{'2}\\
&\vdots\\
&(\underline{L}-\alpha)^{K_c-2}\mathbf{Q_\theta}+(\underline{L}-\alpha)^{K_c}\mathbf{Z}_{L'1}^{'2}+\cdots+(\underline{L}-\alpha)^{K_c+T-1}\mathbf{Z}_{LT}^{'2}\\
&\begin{tikzpicture}\draw[dashed] (0,0)--(9,0);\end{tikzpicture}\\
&\vdots\\
&\begin{tikzpicture}\draw[dashed] (0,0)--(9,0);\end{tikzpicture}\\
&\mathbf{Q_\theta}+(\underline{1}-\alpha)^{K_c}\mathbf{Z}_{11}^{'K_c}+\cdots+(\underline{1}-\alpha)^{K_c+T-1}\mathbf{Z}_{1T}^{'K_c}\\
&\mathbf{Q_\theta}+(\underline{2}-\alpha)^{K_c}\mathbf{Z}_{21}^{'K_c}+\cdots+(\underline{2}-\alpha)^{K_c+T-1}\mathbf{Z}_{2T}^{'K_c}\\
&\vdots\\
&\mathbf{Q_\theta}+(\underline{L}-\alpha)^{K_c}\mathbf{Z}_{f_{L'}1}^{'K_c}+\cdots+(\underline{L}-\alpha)^{K_c+T-1}\mathbf{Z}_{LT}^{'K_c}\\\hline
\multicolumn{2}{c}{\mbox{Desired symbols appear along vectors}}\\
\multicolumn{2}{c}{\vect{(\underline{1}-\alpha)^{-1}},\cdots,\vect{(\underline{L}-\alpha)^{-1}}}\\
\multicolumn{2}{c}{\begin{tikzpicture}\draw[dashed] (0,0)--(9,0);\end{tikzpicture}}\\
\multicolumn{2}{c}{\emphv{\vect{(\underline{1}-\alpha)^{-2}}},\cdots,\emphv{\vect{(\underline{L}-\alpha)^{-2}}},\vect{(\underline{1}-\alpha)^{-1}},\cdots,\vect{(\underline{L}-\alpha)^{-1}}}\\
\multicolumn{2}{c}{\begin{tikzpicture}\draw[dashed] (0,0)--(9,0);\end{tikzpicture}}\\
\multicolumn{2}{c}{\vdots}\\
\multicolumn{2}{c}{\begin{tikzpicture}\draw[dashed] (0,0)--(9,0);\end{tikzpicture}}\\
\multicolumn{2}{c}{\emphv{\vect{(\underline{1}-\alpha)^{-K_c}}},\cdots,\emphv{\vect{(\underline{L}-\alpha)^{-K_c}}}, \cdots,}\\
\multicolumn{2}{c}{\emphv{\vect{(\underline{1}-\alpha)^{-2}}},\cdots,\emphv{\vect{(\underline{L}-\alpha)^{-2}}},}\\
\multicolumn{2}{c}{\vect{(\underline{1}-\alpha)^{-1}},\cdots,\vect{(\underline{L}-\alpha)^{-1}}}\\
\multicolumn{2}{c}{\mbox{Interference  appears along vectors}}\\
\multicolumn{2}{c}{\vect{1}, \vect{(\underline{1}-\alpha)}, \cdots, \vect{(\underline{1}-\alpha)^{X+T+K_c-2}},\cdots,\vect{(\underline{L}-\alpha)}, \cdots, \vect{(\underline{L}-\alpha)^{X+T+K_c-2}}} \\
\multicolumn{2}{c}{\equiv \vect{1},\vect{\alpha},\cdots,\vect{\alpha^{X+T+K_c-2}}}\\\hline
\end{array}
\end{align}
\caption{\it\small A summary of the general MDS-XSTPIR scheme showing storage at each server, the queries, and a partitioning of signal and interference dimensions contained in the answers from each server.}\label{table:summarygen}
\end{table}

\subsection{Arbitrary $U$, $B$}
Now let us generalize the scheme to non-trivial $U$ and $B$, i.e., for  $U$ unresponsive servers and up to $B$ byzantine servers. For this generalization, let us set 
\begin{align}
L&=(N-U)-(K_c+X+T+2B-1)\\
\ell&=LK_c.
\end{align}
Even though now the values of $U, B$ are non-trivial, the construction of storage, queries and answers remains identical to the  description provided previously for $U=B=0$. So let us consider any $(N-U)$ responsive servers, say servers $n_1, n_2, \cdots, n_{N-U}$. Instead of the $N\times N$ square matrix ${\bf M}_{L,N}$ in \eqref{eq:MLNgen}, we now have the $(N-U)\times (N-U-2B)$ decoding matrix,
\begin{align}
\mathbf{M}_{(N-U)\times (N-U-2B)}=\begin{bmatrix}
\frac{1}{\underline{1}-\alpha_{n_1}}&\cdots&\frac{1}{\underline{L}-\alpha_{n_1}}&1&\alpha_{n_1}&\cdots&\alpha_{n_1}^{K_c+X+T-2}\\
\frac{1}{\underline{1}-\alpha_{n_2}}&\cdots&\frac{1}{\underline{L}-\alpha_{n_2}}&1&\alpha_{n_2}&\cdots&\alpha_{n_2}^{K_c+X+T-2}\\
\vdots&\vdots&\vdots&\vdots&\vdots&\vdots&\vdots&\\
\frac{1}{\underline{1}-\alpha_{n_{N-U}}}&\cdots&\frac{1}{\underline{L}-\alpha_{n_{N-U}}}&1&\alpha_{n_{N-U}}&\cdots&\alpha_{n_{N-U}}^{K_c+X+T-2}\\
\end{bmatrix}.
\end{align}
Note that if we consider any $N-U-2B$ rows of $\mathbf{M}_{(N-U)\times (N-U-2B)}$ then we obtain an invertible square matrix because of Lemma \ref{lemma:csa}. Therefore, $\mathbf{M}_{(N-U)\times (N-U-2B)}$ 
 is the generator matrix of an MDS$(N-U,N-U-2B)$ code, and it is can correct up to $((N-U)-(N-U-2B))/2=B$ errors. Thus by this construction, we establish a scheme that works with $U$ unresponsive servers and up to $B$ Byzantine servers, while achieving the rate of
\begin{equation}
R=1-\left(\frac{K_c+X+T+2B-1}{N-U}\right).
\end{equation}
This completes the proof of Theorem \ref{thm:lobound}.

\section{Private and Secure Distributed Matrix Multiplication}\label{sec:PSDMM}
Recently in \cite{Kim_Lee_PSCC, Chang_Tandon_PSDMM}, the problem of  private and secure matrix multiplication (PSDMM) is proposed, where a user wishes to compute the product of a confidential matrix $\mathbf{A}$ with a matrix $\mathbf{B}_{\theta}, \theta\in[M]$ with the aid of $N$ distributed servers. In \cite{Chang_Tandon_PSDMM}, it is assumed that the set of matrices $\mathbf{B}_{[M]}$ are public and available to the $N$ servers, however, the confidential matrix $\mathbf{A}$ is  shared secretly among all $N$ servers, such that no information about $\mathbf{A}$ is leaked to any server. Besides, the user wants to keep the index $\theta$ private from each server. The goal of the problem is to minimize (i) the upload cost from the source of the confidential matrix $\mathbf{A}$ to the $N$ servers and (ii) the download cost from the $N$ servers to the user. In \cite{Chang_Tandon_PSDMM}, the authors exploit the MDS-PIR scheme proposed in \cite{Banawan_Ulukus} to construct the PSDMM scheme, and characterize the lower convex hull of (upload, download) pairs.

Using the MDS-XSTPIR scheme present in Section \ref{sec:achi}, we now present a novel PSDMM scheme for a generalized model. In our model, the index $\theta$ is $T$-private, while  the confidential matrix $\mathbf{A}$ is  $X_A$-secure. Furthermore, we also allow  matrices $\mathbf{B}_{[M]}$ to be $X_B$-secure. Note that the model  in \cite{Chang_Tandon_PSDMM} is obtained as a special case of our generalized model by setting $X_A=T=1, X_B=0$. 

\subsection{PSDMM: Problem Statement}
Let $\mathbf{A}=(\mathbf{A}_1,\mathbf{A}_2,\dots,\mathbf{A}_\ell)$ represent $\ell$ random matrices, each of dimension $\lambda\times\chi$, that are independently and uniformly distributed over  $\mathbb{F}_q^{\lambda\times \chi}$. Let $\mathbf{B}_{[M]}$ be $M$ random matrices independently and uniformly distributed over   $\mathbb{F}_q^{\chi\times \mu}$. The independence between matrices $\mathbf{A}_{[\ell]}$ and $\mathbf{B}_{[M]}$ is formalized as follows.
\begin{equation}
H(\mathbf{A},\mathbf{B}_{[M]})=\sum_{l\in[\ell]}H(\mathbf{A}_l)+\sum_{m\in[M]}H(\mathbf{B}_m).
\end{equation}
The matrices $\mathbf{A}$ and $\mathbf{B}_{[M]}$ are made available at  $N$ distributed servers through  secret sharing schemes with security levels $X_A$ and $X_B$, respectively. That is, any group of up to $X_A$ colluding servers can learn nothing about $\mathbf{A}$, and any group of up to $X_B$ servers can learn nothing about $\mathbf{B}_{[M]}$. To this end, matrices $\mathbf{A}$ and $\mathbf{B}_{[M]}$ are separately coded according to secret sharing schemes that generate shares $\tilde{A}_n$, $\tilde{B}_n, n\in[N]$, and  these shares are made available to the $n$-th server. Furthermore, we assume that the upload cost of $\tilde{A}_{[N]}$ is to be optimized, while that of $\tilde{B}_{[N]}$ and $Q^\theta_{[N]}$ is ignored, presumably because ${\bf A}$ matrices are frequently updated while ${\bf B}_{[M]}$ are static, and the size of queries does not scale with $\ell$.

\begin{figure}[h]
\centerline{\includegraphics[width=5in]{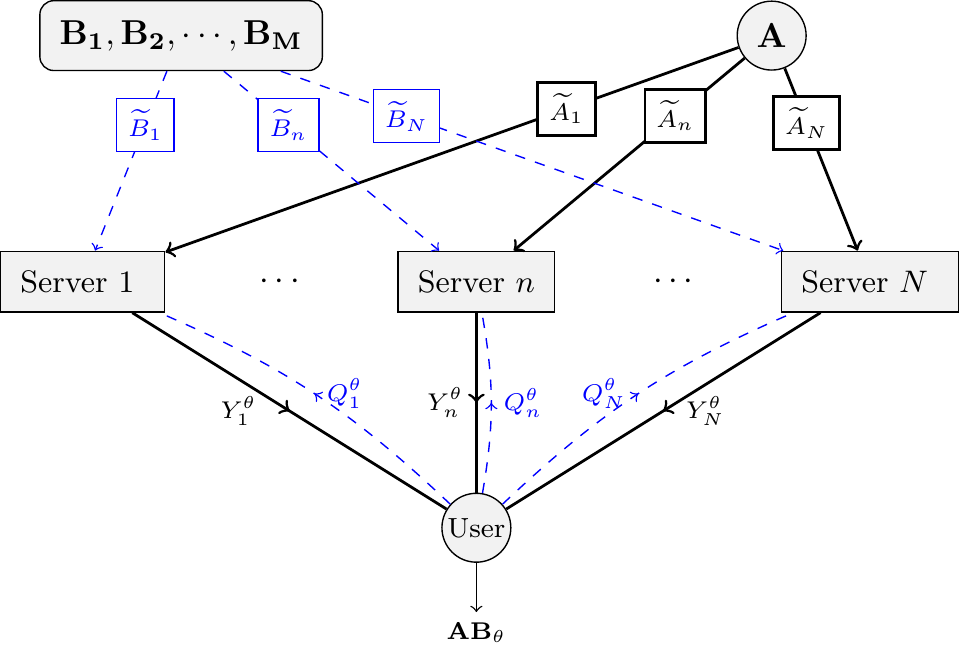}}
\caption{\it \small Model for private secure distributed matrix multiplication (PSDMM). ${\bf A}$ matrices are $X_A$ secure, while ${\bf B}$ matrices are $X_B$ secure. The uploads to be optimized are the $\widetilde{A}$ terms and the downloads to be optimized are the $Y^\theta$ terms.}
\end{figure}

The independence between the securely coded matrices is specified as follows.
\begin{align}
I(\mathbf{A},\tilde{A}_{[N]};\mathbf{B}_{[M]},\tilde{B}_{[N]})=0.
\end{align}
Matrices must be recoverable from their secret shares.
\begin{align}
H(\mathbf{A}\mid \tilde{A}_{[N]})&=0,\\
H(\mathbf{B}_{[M]}\mid \tilde{B}_{[N]})&=0.
\end{align}
The matrices must be perfectly secure from any set of secret shares that can be accessed by a set of up to $X_A, X_B$ colluding servers, respectively.
\begin{align}
I(\mathbf{A};\tilde{A}_{\mathcal{X}})&=0 \quad \mathcal{X}\subset[N], |\mathcal{X}|=X_A,\\
I(\mathbf{B}_{[M]};\tilde{B}_{\mathcal{X}})&=0 \quad \mathcal{X}\subset[N], |\mathcal{X}|=X_B.
\end{align}
The user generates an index $\theta\in[M]$ privately and uniformly, and wishes to compute the product
\begin{equation}
\mathbf{A}\mathbf{B}_{\theta}=(\mathbf{A}_1\mathbf{B}_{\theta},\mathbf{A}_2\mathbf{B}_{\theta},\dots,\mathbf{A}_\ell\mathbf{B}_{\theta}).
\end{equation}
To this end, the user generates $N$ queries $Q_{[N]}^{\theta}$. The $n$-th query $Q_n^{\theta}$ is sent to the $n$-th server. The user has no prior knowledge of matrices $\mathbf{A}$ and $\mathbf{B}_{[M]}$ and their secret shares, i.e.,
\begin{equation}
I(\theta,Q_{[N]}^{\theta};\tilde{A}_{[N]},\tilde{B}_{[N]})=0.
\end{equation}
$T$-privacy, $0\leq T\leq N$, guarantees that any group of up to $T$ colluding servers learn nothing about $\theta$.
\begin{equation}
I(Q_{\mathcal{T}}^{\theta},\tilde{A}_{\mathcal{T}},\tilde{B}_{\mathcal{T}};\theta)=0.
\end{equation}
Upon receiving the user's query $Q_n^{\theta}$, the $n$-th server responds  with an answer $Y_n^{\theta}$, which is a function of all information available to it.
\begin{equation}
H(Y_n^{\theta}|Q_n^{\theta},\tilde{A}_{n},\tilde{B}_{n})=0.
\end{equation}
The user must be able to recover the product $\mathbf{A}\mathbf{B}_{\theta}$ from all $N$ answers, i.e.,
\begin{equation}
H(\mathbf{A}\mathbf{B}_{\theta}|Y_{[N]}^{\theta},Q_{[N]}^{\theta})=0.
\end{equation}
The upload cost and download cost are defined as follows.
\begin{align}
U&=\frac{\sum_{n\in{[N]}}H(\tilde{A}_n)}{H(\mathbf{A})},\\
D&=\frac{\sum_{n\in{[N]}}H(Y_n^{\theta})}{H(\mathbf{A}\mathbf{B}_{\theta})}.
\end{align}

\subsection{A New Scheme for PSDMM}
In this section, we will present a  PSDMM scheme to show that the lower convex hull of (upload, download) pairs
\begin{equation}
(U,D)=\left(\frac{N}{K_c},~~\frac{N}{N-(2K_c+X_A+X_B+T-2)}\right)
\end{equation}
for 
\begin{align}
K_c&=1,2,\dots,\lfloor(N+1-X_A-X_B-T)/2\rfloor
\end{align}
 is achievable when $q\rightarrow\infty$ and $\chi\geq\min(\lambda,\mu)$. Furthermore, when $X_B=0$, i.e., there are no security constraints on matrices $\mathbf{B}_{[M]}$, and $\chi\geq\min(\lambda,\mu)$, then the lower convex hull of (upload, download) pairs
\begin{equation}
(U,D)=\left(\frac{N}{K_c},\frac{N}{N-(K_c+X_A+T-1)}\right)
\end{equation}
for 
\begin{align}
K_c&=1,2,\dots,(N+1-X_A-T)
\end{align} is achievable as $q\rightarrow\infty$.

First, let us consider the case $X_B\neq 0$. For this setting, let us set
\begin{align}
L&=N-(X_A+X_B+T+2K_c-2),\\
\ell&=K_cL.
\end{align}
For all $l\in[L], k\in[K_c]$, let us define
\begin{equation}
\mathbf{A}_{l k}=\mathbf{A}_{L(k-1)+l}.
\end{equation}
We will also set
\begin{equation}
\mathbf{B}=\begin{bmatrix}
\mathbf{B}_1&\mathbf{B}_2&\dots&\mathbf{B}_M
\end{bmatrix}
\end{equation}
to be an $\chi\times M\mu$ matrix that contains all $\mathbf{B}_{[M]}$. Let us also define $\mathbf{Q}_{\theta}$ be a $M\mu\times \mu$ matrix as follows.
\begin{equation}
\mathbf{Q}_{\theta}=[
\underbrace{\mathbf{0}_{\mu}\quad\dots\quad\mathbf{0}_{\mu}}_{\text{A total of $(\theta-1) \mathbf{0}_{\mu}$'s}}\quad\mathbf{I}_{\mu}\quad\underbrace{\mathbf{0}_{\mu}\quad\dots\quad\mathbf{0}_{\mu}}_{\text{A total of $(M-\theta) \mathbf{0}_{\mu}$'s}}]^T
\end{equation}
where $\mathbf{0}_{\mu}$ is the ${\mu}\times {\mu}$ square zero matrix, and $\mathbf{I}_{\mu}$ is the ${\mu}\times {\mu}$ identity matrix. We note that by  construction, $\mathbf{A}\mathbf{B}\mathbf{Q}_{\theta}=(\mathbf{A}_1\mathbf{B}\mathbf{Q}_{\theta},\dots,\mathbf{A}_\ell\mathbf{B}\mathbf{Q}_{\theta})=(\mathbf{A}_{l k}\mathbf{B}\mathbf{Q}_{\theta})_{l\in[L],k
\in[K_c]}$ is the desired product. Let $(\mathbf{Z}_{l x})_{l\in[L], x\in[X_A]}$ and $(\mathbf{Z}'_{l x'})_{l\in[L], x'\in[X_B]}$ be independent, uniformly distributed random noise matrices from $\mathbb{F}_q^{\lambda\times \chi}$ and $\mathbb{F}_q^{\chi\times M\mu}$ that will be used to guarantee $X_A$ and $X_B$ security levels for $\mathbf{A}, \mathbf{B}_{[M]}$, respectively. Let $(\mathbf{Z}_{l t}^{''\kappa})_{l\in[L], t\in[T], \kappa\in[K_c]}$ be independent, uniformly distributed random noise matrices from $\mathbb{F}_q^{M\mu\times \mu}$, that will be used to guarantee $T$-privacy of queries. The independence between $\mathbf{A}, \mathbf{B}_{[M]}$, noise matrices and $\theta$ is formalized as follows.
\begin{align}
&H(\mathbf{A}, \mathbf{B}_{[M]}, (\mathbf{Z}_{l x})_{l\in[L], x\in[X_A]}, (\mathbf{Z}'_{l x'})_{l\in[L], x'\in[X_B]}, (\mathbf{Z}_{l t}^{''\kappa})_{l\in[L], t\in[T], \kappa\in[K_c]}, \theta)\notag\\
&=\sum_{l\in[L],k\in[K_c]}H(\mathbf{A}_{l k})+\sum_{m\in[M]}H(\mathbf{B}_m)+\sum_{l\in[L], x\in[X_A]}H(\mathbf{Z}_{l x})\notag\\
&\quad\quad+\sum_{l\in[L], x'\in[X_B]}H(\mathbf{Z}'_{l x'})+\sum_{l\in[L], t\in[T], \kappa\in[K_c]}H(\mathbf{Z}_{l t}^{''\kappa})+H(\theta).
\end{align}

Let $\underline{1}, \underline{2},\cdots,\underline{L}, \alpha_1,\alpha_2,\cdots, \alpha_N$ be distinct elements of $\mathbb{F}_q$. We require $q\geq L+N$ so these elements must exist. Now we are ready to construct the scheme. The secret share of $\mathbf{B}_{[M]}$ at the $n$-th server, $\tilde{B}_n$ is constructed as follows.
\begin{equation}
\tilde{B}_n=(\tilde{B}_{n1}, \tilde{B}_{n2}, \dots, \tilde{B}_{nL}),
\end{equation}
where $\forall l\in[L]$,
\begin{equation}
\tilde{B}_{nl}=\mathbf{B}+\sum_{x'\in[X_B]}(\underline{l}-\alpha_n)^{K_c+x'-1}\mathbf{Z}'_{l x'}.
\end{equation}
The secret share of $\mathbf{A}$ at the $n^{th}$ server is constructed as follows.
\begin{equation}
\tilde{A}_n=(\tilde{A}_{n1}, \tilde{A}_{n2}, \dots, \tilde{A}_{nL}),
\end{equation}
where $\forall l\in[L]$,
\begin{equation}
\tilde{A}_{nl}=\sum_{k\in[K_c]}\frac{1}{(\underline{l}-\alpha_n)^{K_c-k+1}}\mathbf{A}_{l k}+\sum_{x\in[X_A]}(\underline{l}-\alpha_n)^{x-1}\mathbf{Z}_{l x}.
\end{equation}
The query sent by the user to the $n^{th}$ server, is comprised of $K_c$ rounds, $Q_n^\theta=(Q_n^{\theta, \kappa})_{\kappa\in[K_c]}$. For all $\kappa\in[K_c]$, we construct the queries as follows.
\begin{equation}
Q_n^{\theta,\kappa}=(Q_{n1}^{\theta,\kappa},Q_{n2}^{\theta,\kappa},\dots,Q_{nL}^{\theta,\kappa}),
\end{equation}
where $\forall l\in[L]$, we set
\begin{equation}
Q_{nl}^{\theta,\kappa}=(\underline{l}-\alpha_n)^{K_c-\kappa}\mathbf{Q}_{\theta}+\sum_{t\in[T]}(\underline{l}-\alpha_n)^{K_c+t-1}\mathbf{Z}_{l t}^{''\kappa}.
\end{equation}
Upon receiving the query from the user, the $n^{th}$ server responds with the following $K_c$ symbols.
\begin{equation}
Y_n^{\theta}=(\tilde{A}_{n1}\tilde{B}_{n1}Q_{n1}^{\theta,\kappa}+\tilde{A}_{n2}\tilde{B}_{n2}Q_{n2}^{\theta,\kappa}+\dots+\tilde{A}_{nL}\tilde{B}_{nL}Q_{nL}^{\theta,\kappa})_{\kappa\in[K_c]}.
\end{equation}
To show the correctness of the scheme, let us consider $\tilde{A}_{nl}\tilde{B}_{nl}, \forall l\in[L]$.
\begin{align}
\tilde{A}_{nl}\tilde{B}_{nl}&=\left(\sum_{k\in[K_c]}\frac{1}{(\underline{l}-\alpha_n)^{K_c-k+1}}\mathbf{A}_{l k}+\sum_{x\in[X_A]}(\underline{l}-\alpha_n)^{x-1}\mathbf{Z}_{l x}\right)\notag\\
&\quad\quad\quad\quad\left(\mathbf{B}+\sum_{x'\in[X_B]}(\underline{l}-\alpha_n)^{K_c+x'-1}\mathbf{Z}'_{l x'}\right)\\
&=\sum_{k\in[K_c]}\frac{1}{(\underline{l}-\alpha_n)^{K_c-k+1}}\mathbf{A}_{l k}\mathbf{B}+\sum_{x\in[X_A]}(\underline{l}-\alpha_n)^{x-1}\mathbf{Z}_{l x}\mathbf{B}\notag\\
&\quad\quad+\sum_{k\in[K_c]}\sum_{x'\in[X_B]}(\underline{l}-\alpha_n)^{x'+k-2}\mathbf{A}_{lk}\mathbf{Z}'_{\ell x'}+\sum_{x\in[X_A]}\sum_{x'\in[X_B]}(\underline{l}-\alpha_n)^{K_c+x+x'-2}\mathbf{Z}_{l x}\mathbf{Z}'_{l x'}\label{eq:penstep}\\
&=\sum_{k\in[K_c]}\frac{1}{(\underline{l}-\alpha_n)^{K_c-k+1}}\mathbf{A}_{l k}\mathbf{B}+\sum_{\xi\in[K_c+X_A+X_B-1]}(\underline{l}-\alpha_n)^{\xi-1}\bar{\bf Z}_{l\xi}\label{eq:laststep}
\end{align}
In \eqref{eq:laststep} we rearranged the last three terms of \eqref{eq:penstep} grouping them into weighted sums of terms of the form $(\underline{l}-\alpha_n)^i$, $i\in\{0,1,\dots,K_c+X_A+X_B-2\}$. The grouped terms $\bar{\bf Z}_{l\xi}$ can be calculated explicitly but as it turns out the precise form of these terms is inconsequential. Now note that if we regard $(\mathbf{A}_{l k}\mathbf{B})_{l\in[L],k\in[K_c]}$ terms as messages, and other terms as noise, then \eqref{eq:laststep} has the same form as \eqref{eq:schemestor}, the storage construction in the MDS-XSTPIR scheme presented in Section \ref{sec:achi}.\footnote{Note that $X$ in \eqref{eq:schemestor} corresponds to $K_c+X_A+X_B-1$ in \eqref{eq:laststep}, so that $L=N-(X+T+K_c-1)$ in Section \ref{sec:achi} corresponds to $L=N-(2K_c+X_A+X_B+T-2)$ in this section. The condition on $K_c$ becomes $K_c=\frac{N-(X_A+X_B+T+L-2)}{2}$. However, since we must have $L\geq 1$ and $K_c\geq 1$ can only take integer values, it follows that the feasible values of $K_c$ are $1\leq K_c\leq \lfloor\frac{N-(X_A+X_B+T-1)}{2}\rfloor$.} Also note that the construction of queries is also the same as the MDS-XSTPIR scheme, thus the correctness follows directly from the proof presented in Section \ref{sec:achi}, which means the user is able to recover the product $\mathbf{A}\mathbf{B}\mathbf{Q}_{\theta}=(\mathbf{A}_{l k}\mathbf{B}\mathbf{Q}_{\theta})_{l\in[L],k\in[K_c]}$. Privacy and security follows from the fact that $\mathbf{Q}_{\theta}$, $\mathbf{A}$, $\mathbf{B}_{[M]}$ are protected by the i.i.d. uniformly distributed noise matrices coded according to MDS$(N,T)$, MDS$(X_A,T)$, MDS$(X_B,T)$ codes, respectively. This completes the construction of the scheme for $X_B\neq 0$. Note that when $q\rightarrow\infty$ and $\chi\geq\min(\lambda,\mu)$, then $H(\mathbf{A}\mathbf{B}_{\theta})=\ell\lambda\mu$ in $q$-ary units according to (\cite{Jia_Jafar_SDMM}, Lemma 2), and the download cost is
\begin{equation}
D=\frac{NK_c\lambda\mu}{\ell\lambda\mu}=\frac{N}{L}=\frac{N}{N-(2K_c+X_A+X_B+T-2)}.
\end{equation}

Now let us consider the case $X_B=0$. For this setting, let us set
\begin{align}
L&=N-(X_A+X_B+T+K_c-1),\\
\ell&=K_cL.
\end{align}
We will continue using other definitions as before, but since there is no security constraint on ${\bf B}$ matrices, let us replace $\tilde{B}_n$ as
\begin{equation}
\tilde{B}_n=\mathbf{B}.
\end{equation}
Now we have
\begin{align}
\tilde{A}_{nl}\tilde{B}_{nl}
&=\sum_{k\in[K_c]}\frac{1}{(\underline{l}-\alpha_n)^{K_c-k+1}}\mathbf{A}_{l k}\mathbf{B}+\sum_{x\in[X_A]}(\underline{l}-\alpha_n)^{x-1}\mathbf{Z}_{l x}\mathbf{B},
\end{align}
which is coded according to an MDS$(N, K_c+X_A)$ code. Thus the correctness, privacy and security follows from that proof in Section \ref{sec:achi}. The download cost is
\begin{equation}
D=\frac{NK_c\lambda\mu}{L\lambda\mu}=\frac{N}{L}=\frac{N}{N-(K_c+X_A+X_B+T-1)}.
\end{equation}
Now let us consider the upload cost of the scheme. Note that by the construction of $\tilde{A}_n$, it is coded according to an MDS$(N,K_c)$ code. Therefore, the upload cost is $\frac{N}{K_c}$. 

It is shown in \cite{Chang_Tandon_PSDMM} that when $X_A=T=1, X_B=0$, the lower convex hull of (upload, download) pairs
\begin{equation}
(U,D)=\left(\frac{N}{K_c}, \frac{K_c+1}{K_c}\left(1+\left(\frac{K_c+1}{N}\right)+\dots+\left(\frac{K_c+1}{N}\right)^{M-1}\right)\right)
\end{equation}
is achievable for $K_c=1,2,\dots,N-1$. For the asymptotic setting, i.e., $M\rightarrow\infty$, we have from \cite{Chang_Tandon_PSDMM} that  $D=\frac{K_c+1}{K_c}\frac{N}{N-(K_c+1)}$, which is strictly worse than the $(\mbox{upload},\mbox{download})$ pairs characterized in this work. This is because the scheme in \cite{Chang_Tandon_PSDMM} allows the user to decode noise matrices protecting $\mathbf{A}$, whereas in our scheme, because of \emph{cross-subspace alignment}, the user is only able to decode desired matrices, thus the penalty term $\frac{K_c+1}{K_c}$ disappears.

\section{Conclusion}
The problem of U-B-MDS-XSTPIR, i.e., $X$-secure $T$-private information retrieval from MDS coded storage, with $N$ servers out of which $U$ are unresponsive and up to $B$ may be Byzantine,  is studied in this work. A lower bound on achievable rates of U-B-MDS-XSTPIR is characterized by presenting a \emph{cross-subspace alignment} and successive decoding based scheme. We also adapt the scheme to the problem of private and secure distributed matrix multiplication that is recently proposed in \cite{Kim_Lee_PSCC, Chang_Tandon_PSDMM}. The presented MDS-XSTPIR scheme is shown to be applicable to PSDMM problem, even if we allow security concerns for all constituent matrices. The immediate challenge for future work is to settle the asymptotic capacity conjectures for MDS-TPIR, and also of MDS-XSTPIR, either in the affirmative by finding tight converse bounds or in the negative by finding better asymptotic achievable schemes. Beyond this, settling down the conjecture of asymptotic capacity of U-B-MDS-XSTPIR with unresponsive and Byzantine servers also merits  investigation.

\bibliographystyle{IEEEtran}
\bibliography{Thesis}

\begin{thebibliography}{10}
\providecommand{\url}[1]{#1}
\csname url@samestyle\endcsname
\providecommand{\newblock}{\relax}
\providecommand{\bibinfo}[2]{#2}
\providecommand{\BIBentrySTDinterwordspacing}{\spaceskip=0pt\relax}
\providecommand{\BIBentryALTinterwordstretchfactor}{4}
\providecommand{\BIBentryALTinterwordspacing}{\spaceskip=\fontdimen2\font plus
\BIBentryALTinterwordstretchfactor\fontdimen3\font minus
  \fontdimen4\font\relax}
\providecommand{\BIBforeignlanguage}[2]{{%
\expandafter\ifx\csname l@#1\endcsname\relax
\typeout{** WARNING: IEEEtran.bst: No hyphenation pattern has been}%
\typeout{** loaded for the language `#1'. Using the pattern for}%
\typeout{** the default language instead.}%
\else
\language=\csname l@#1\endcsname
\fi
#2}}
\providecommand{\BIBdecl}{\relax}
\BIBdecl

\bibitem{PIRfirst}
B.~Chor, O.~Goldreich, E.~Kushilevitz, and M.~Sudan, ``Private information
  retrieval,'' in \emph{Proceedings of the 36th Annual Symposium on Foundations
  of Computer Science}, 1995, pp. 41--50.

\bibitem{Sun_Jafar_PIR}
H.~Sun and S.~A. Jafar, ``{The Capacity of Private Information Retrieval},''
  \emph{IEEE Transactions on Information Theory}, vol.~63, no.~7, pp.
  4075--4088, July 2017.

\bibitem{Tajeddine_Rouayheb}
R.~Tajeddine, O.~W. Gnilke, and S.~El~Rouayheb, ``{Private Information
  Retrieval from {MDS} Coded Data in Distributed Storage Systems},'' \emph{IEEE
  Transactions on Information Theory}, 2018.

\bibitem{LDC}
J.~Katz and L.~Trevisan, ``On the efficiency of local decoding procedures for
  error-correcting codes,'' in \emph{Proceedings of the thirty-second annual
  ACM symposium on Theory of computing}.\hskip 1em plus 0.5em minus 0.4em\relax
  ACM, 2000, pp. 80--86.

\bibitem{YekhaninPhd}
S.~Yekhanin, ``{Locally Decodable Codes and Private Information Retrieval
  Schemes},'' Ph.D. dissertation, Massachusetts Institute of Technology, 2007.

\bibitem{Gopalan_Huang_Simitci_Yekhanin}
P.~Gopalan, C.Huang, H.~Simitci, and S.~Yekhanin, ``{On the Locality of
  Codeword Symbols},'' \emph{IEEE Transactions on Information Theory}, vol.~58,
  no.~11, pp. 6925--6934, Nov. 2012.

\bibitem{Batch}
Y.~Ishai, E.~Kushilevitz, R.~Ostrovsky, and A.~Sahai, ``{Batch codes and their
  applications},'' in \emph{Proceedings of the thirty-sixth annual ACM
  symposium on Theory of computing}.\hskip 1em plus 0.5em minus 0.4em\relax
  ACM, 2004, pp. 262--271.

\bibitem{Rabin}
M.~O. Rabin, ``How to exchange secrets with oblivious transfer.'' 1981.

\bibitem{SymPIR}
Y.~Gertner, Y.~Ishai, E.~Kushilevitz, and T.~Malkin, ``Protecting data privacy
  in private information retrieval schemes,'' in \emph{Proceedings of the
  thirtieth annual ACM symposium on Theory of computing}.\hskip 1em plus 0.5em
  minus 0.4em\relax ACM, 1998, pp. 151--160.

\bibitem{Hide_one}
M.~Abadi, J.~Feigenbaum, and J.~Kilian, ``{On hiding information from an
  oracle},'' in \emph{Proceedings of the nineteenth annual ACM symposium on
  Theory of computing}.\hskip 1em plus 0.5em minus 0.4em\relax ACM, 1987, pp.
  195--203.

\bibitem{Shamir}
A.~Shamir, ``How to share a secret,'' \emph{Communications of the ACM},
  vol.~22, pp. 612--613, 1979.

\bibitem{Jafar_corr}
S.~A. Jafar, ``{Blind Interference Alignment},'' \emph{IEEE Journal of Selected
  Topics in Signal Processing}, vol.~6, no.~3, pp. 216--227, June 2012.

\bibitem{Sun_Jafar_BIAPIR}
H.~Sun and S.~A. Jafar, ``Blind interference alignment for private information
  retrieval,'' \emph{2016 IEEE International Symposium on Information Theory
  (ISIT)}, pp. 560--564, 2016.

\bibitem{yao1982protocols}
A.~C. Yao, ``Protocols for secure computations,'' in \emph{Foundations of
  Computer Science, 1982. SFCS'08. 23rd Annual Symposium on}.\hskip 1em plus
  0.5em minus 0.4em\relax IEEE, 1982, pp. 160--164.

\bibitem{Chang_Tandon_SDMMOS}
W.-T. Chang and R.~Tandon, ``On the capacity of secure distributed matrix
  multiplication,'' \emph{arXiv preprint arXiv:1806.00469}, 2018.

\bibitem{DOliveira_Rouayheb_Karpuk}
R.~G. D'Oliveira, S.~E. Rouayheb, and D.~Karpuk, ``Gasp codes for secure
  distributed matrix multiplication,'' \emph{arXiv preprint arXiv:1812.09962},
  2018.

\bibitem{Kakar_Ebadifar_Sezgin}
J.~Kakar, S.~Ebadifar, and A.~Sezgin, ``Rate-efficiency and
  straggler-robustness through partition in distributed two-sided secure matrix
  computation,'' \emph{arXiv preprint arXiv:1810.13006}, 2018.

\bibitem{Aliasgari_Simeone_Kliewer}
M.~Aliasgari, O.~Simeone, and J.~Kliewer, ``Distributed and private coded
  matrix computation with flexible communication load,'' \emph{arXiv preprint
  arXiv:1901.07705}, 2019.

\bibitem{Jia_Jafar_SDMM}
Z.~Jia and S.~Jafar, ``On the capacity of secure distributed matrix
  multiplication,'' \emph{arXiv preprint arXiv:1908.06957}, 2019.

\bibitem{Chang_Tandon_PSDMM}
W.~Chang and R.~Tandon, ``On the upload versus download cost for secure and
  private matrix multiplication,'' \emph{arXiv preprint arXiv:1906.10684},
  2019.

\bibitem{Sun_Jafar_TPIR}
H.~Sun and S.~A. Jafar, ``{The Capacity of Robust Private Information Retrieval
  with Colluding Databases},'' \emph{IEEE Transactions on Information Theory},
  vol.~64, no.~4, pp. 2361--2370, April 2018.

\bibitem{Banawan_Ulukus_BPIR}
K.~{Banawan} and S.~{Ulukus}, ``The capacity of private information retrieval
  from byzantine and colluding databases,'' \emph{IEEE Transactions on
  Information Theory}, vol.~65, no.~2, pp. 1206--1219, Feb 2019.

\bibitem{Tajeddine_Gnilke_Karpuk_Hollanti}
R.~{Tajeddine}, O.~W. {Gnilke}, D.~{Karpuk}, R.~{Freij-Hollanti}, and
  C.~{Hollanti}, ``Private information retrieval from coded storage systems
  with colluding, byzantine, and unresponsive servers,'' \emph{IEEE
  Transactions on Information Theory}, vol.~65, no.~6, pp. 3898--3906, June
  2019.

\bibitem{Banawan_Ulukus}
K.~Banawan and S.~Ulukus, ``{The Capacity of Private Information Retrieval from
  Coded Databases},'' \emph{IEEE Transactions on Information Theory}, vol.~64,
  no.~3, pp. 1945--1956, 2018.

\bibitem{Jia_Sun_Jafar_XSTPIR}
Z.~{Jia}, H.~{Sun}, and S.~A. {Jafar}, ``Cross subspace alignment and the
  asymptotic capacity of $x$ -secure $t$ -private information retrieval,''
  \emph{IEEE Transactions on Information Theory}, vol.~65, no.~9, pp.
  5783--5798, Sep. 2019.

\bibitem{FREIJ_HOLLANTI}
R.~Freij-Hollanti, O.~Gnilke, C.~Hollanti, and D.~Karpuk, ``{Private
  Information Retrieval from Coded Databases with Colluding Servers},''
  \emph{SIAM Journal on Applied Algebra and Geometry}, vol.~1, no.~1, pp.
  647--664, 2017.

\bibitem{Sun_Jafar_MDSTPIR}
H.~Sun and S.~A. Jafar, ``{Private Information Retrieval from MDS Coded Data
  with Colluding Servers: Settling a Conjecture by Freij-Hollanti et al.}''
  \emph{IEEE Transactions on Information Theory}, vol.~64, no.~2, pp.
  1000--1022, February 2018.

\bibitem{Kim_Lee_PSCC}
M.~{Kim} and J.~{Lee}, ``Private secure coded computation,'' \emph{IEEE
  Communications Letters}, pp. 1--1, 2019, doi: 10.1109/LCOMM.2019.2934436.

\bibitem{Shah_Rashmi_Kannan}
N.~Shah, K.~Rashmi, and K.~Ramchandran, ``{One Extra Bit of Download Ensures
  Perfectly Private Information Retrieval},'' in \emph{Proceedings of IEEE
  International Symposium on Information Theory (ISIT)}, 2014, pp. 856--860.

\bibitem{Raviv_Karpuk}
N.~{Raviv} and D.~A. {Karpuk}, ``Private polynomial computation from lagrange
  encoding,'' \emph{IEEE Transactions on Information Forensics and Security},
  pp. 1--1, 2019, doi: 10.1109/TIFS.2019.2925723.

\end{thebibliography}

\end{document}